\documentclass{aastex62}

\newcommand{\degree}{^{\circ}}

\received{2019 November 27}
\revised{2020 January 21}
\accepted{2020 January 23}
\submitjournal{and accepted by the Astronomical Journal}

\shorttitle{Near-Sun comets}
\shortauthors{Wiegert et al.}

\begin{document}

\title{Supercatastrophic disruption of asteroids in the context of SOHO comet, fireball and meteor observations}

\correspondingauthor{Paul Wiegert}
\email{pwiegert@uwo.ca}

\author{Paul Wiegert}
\affil{Department of Physics and Astronomy, The University of Western Ontario, London, Canada}
\affil{Institute for Earth and Space Exploration (IESX), The University of Western Ontario, London, Canada}

\author{Peter Brown}
\affil{Department of Physics and Astronomy, The University of Western Ontario, London, Canada}
\affil{Institute for Earth and Space Exploration (IESX), The University of Western Ontario, London, Canada}

\author{Petr Pokorn\'{y}}
\affil{Department of Physics, The Catholic University of America, Washington, DC 20064, USA}
\affil{Astrophysics Science Division, NASA Goddard Space Flight Center, Greenbelt, MD 20771}

\author{Quanzhi Ye}
\affil{Department of Astronomy, University of Maryland, College Park, MD 20742, USA}

\author{Cole Gregg}
\affil{Department of Physics and Astronomy, The University of Western Ontario, London, Canada}

\author{Karina Lenartowicz}
\affil{Department of Physics and Astronomy, The University of Western Ontario, London, Canada}

\author{Zbigniew Krzeminski}
\affil{Department of Physics and Astronomy, The University of Western Ontario, London, Canada}

\author{David Clark}
\affil{Department of Earth Sciences, The University of Western Ontario, London, Canada}
\affil{Institute for Earth and Space Exploration (IESX), The University of Western Ontario, London, Canada}



\begin{abstract}
\cite{gramorjed16} report an absence of asteroids on orbits with
perihelia near the Sun that they attribute to the 'supercatastrophic
disruption' of these bodies. Here we investigate whether there is
evidence for this process among other bodies with similarly
low perihelia: near-Earth asteroids, SOHO comets, as well as meter-sized and millimeter-sized meteoroids. We determine no known near-Earth asteroids have past (last $10^4$ years) histories residing significantly inside the \cite{gramorjed16} limit, indirectly supporting the disruption hypothesis. The exception is asteroid (467372) 2004 LG which spent 2500 years within this limit, and thus presents a challenge to that theory. Phaethon has a perihelion distance hovering just above the limit and may be undergoing slow disruption, which may be the source of its dust complex. We find that the rate at which ungrouped SOHO comets are observed is consistent with expected rates for the injection of small (25 m) class asteroids into the near-Sun region and suggest that this fraction of the SOHO-observed comet population may in fact be asteroidal in origin. We also find that there is an absence of meter-sized bodies with near-Sun perihelia but an excess of millimeter-sized meteoroids. This implies that if near-Sun asteroids disrupt, they do not simply fragment into meter-sized chunks but disintegrate ultimately into millimeter-sized particles. 
We propose that the disruption of near-Sun asteroids as well as the anomalous brightening and destruction processes that affect SOHO comets occur through meteoroid erosion, that is, the removal of material through impacts by high-speed near-Sun meteoroids.

\end{abstract}



\section{Introduction} \label{intro}
It has been known for some time that near-Earth asteroids are lost largely to orbital evolution resulting in their falling into the Sun \citep{farfrofro94, glamigmor97}. However it has recently been reported that asteroids may be destroyed even when their perihelion is substantially above the solar surface and their temperatures remain below that needed to melt/vaporize them. On the basis of an absence of known near-Earth asteroids with small perihelion, even after painstaking removal of observational selection effects, \cite{gramorjed16} (hereafter G16) concluded that asteroids 'supercatastrophically disrupt' (SCD) when their perihelion $q$ reaches values below about 16 solar radii $R_{\odot}$ (or 0.074 AU), with some dependence on asteroid size. Their analysis determined that neither tidal effects nor thermal melting/vaporization can adequately account for the destruction of these asteroids. No physical mechanism was proposed but G16 speculated that the break-up may be due to thermal cracking, spin-up beyond the asteroids' cohesive strength, or explosive fracturing due to subsurface volatile release.

Here we examine the near-Sun population across a range of sizes to look for clues to the SCD process. The known near-Sun asteroids, in particular their past dynamical history;  the SOHO comets; meter-sized fireballs and millimeter-sized meteors from the near-Sun region all carry information that may be relevant to the SCD process.  
First we note that it is not clear that the SCD process is distinct from the processes of what will refer to here as 'ordinary' cometary activity and/or 'ordinary' cometary fragmentation. Though the process must result in the destruction of asteroids, which may not have much ice content, it is possible that SCD may look very much like ordinary cometary fragmentation. It is even possible that after one or a few perihelion passages near the Sun, inner volatiles are exposed by some process, and the 'asteroid' becomes what is --- for all intents and purposes --- an ordinary comet, even displaying cometary activity when relatively far from the Sun. Thus bodies currently-labelled as 'comets' with low-perihelia (SOHO comets, for example) may in fact be undergoing SCD. In particular, we note that since the \cite{gramorjed18} near-Earth asteroid model includes a Jupiter-Family Comet (JFC) source, it would seem that JFCs likely undergo SCD either by the same or a similar process as asteroids do, because they must be driven to the near-Sun region by similar dynamical processes but do not survive there.  Thus comets are not necessarily immune from SCD. That being said, comets do undergo splittings far from the Sun as well. Over 40 split comets were reviewed by \cite{boe04}. For none except Shoemaker-Levy 9 (which was tidally disrupted by a close approach to Jupiter) is the mechanism well-understood, and some were thought to have split at heliocentric distances beyond 50~AU. Thus while comets may be vulnerable to SCD, it is in addition to other disruption processes.

In the next section, we'll review the known populations of low-perihelion bodies in the context of SCD. 

\section{The near-Sun populations} \label{methods}

\subsection{Near-Sun asteroids}
\label{pastperihelion}
The current distribution of perihelion distances of near-Sun asteroids shows a deficit near the Sun \citep{gramorjed16}, but there are many asteroids with $q$ near this limit that might have been closer to the Sun in the past. That is because the same dynamical effects which drive asteroid perihelia into the near-Sun region can also draw them out again, if the asteroids survive. Thus the current population of asteroids contains some asteroids which were at lower $q$ in the past, and thus potentially provide tests of and information on the SCD phenomenon. 

To perform a preliminary investigation, we select all the asteroids listed on the NeoDys website\footnote{https://newton.spacedys.com/neodys/ retrieved 4 April 2019} that have a perihelion distance $q$ less that 0.5~AU, with absolute magnitude $H<19$ and whose orbit condition code $U$ is greater than or equal to 2 according to JPL\footnote{https://ssd.jpl.nasa.gov/sbdb.cgi, retrieved 5 April 2019}. The quantity $U$ runs from 0 to 9 and reflects how well the orbit is known, with 0 indicating little uncertainty, and 9 indicating high uncertainty\footnote{https://minorplanetcenter.net//iau/info/UValue.html retrieved 7 Jan 2020}. The resulting sample contains the largest NEAs (diameters above 500 m, depending on albedo) at these perihelia, and have the best determined orbits.

The nominal orbits of these bodies are integrated backwards for 10,000 years within a solar system with eight planets (using the Earth-Moon barycenter for the Earth) and post-newtonian relativistic corrections, with the RADAU \citep{eve85} integrator with an error tolerance of $10^{-12}$.  During the integration, passages within the G16 limit for asteroids of this size (0.06~AU) are checked for.  This is not an exhaustive examination of the NEA population: our sample contains only 1324 of the 19771 objects on the NeoDys list, but is intended to capture the best-known orbits most likely to have crossed the G16 limit in the past. A full study of the entire NEA population for past crossings of the G16 limit would be interesting but beyond the scope of this paper.

Though a number of our sample asteroids approach or barely cross the G16 limit, e.g asteroid (511600) 2015 AZ$_{245}$ had $q = 0.058$~AU about 4600 years in the past, the vast majority of asteroids we examined remain outside the limit during the interval examined. This strengthens the case for SCD: if many large asteroids had survived extended periods with perihelia inside the G16 limit, it might point to the current deficit being a statistical fluke.

There is an exception: asteroid (467372) 2004 LG (which has semimajor axis $a=2.06$~AU, eccentricity $e=0.897$, inclination $i=71\degree$) reached $q = 0.026$~AU about 2400 years ago, half the SCD limit at its size ($H=18.0$) and half that of any other asteroid in our sample. It spent approximately 2500~yr within $16 R_{\odot}$.  We repeated this simulation with 100 clones of 467372 generated from its NeoDys covariance matrix and all showed the same behaviour, so this is not simply the result of a poorly-determined orbit. Asteroid (467372) 2004 LG is currently outside the G16 limit at $q=0.21$~AU and an inclination $i = 71\degree$. This highly-inclined orbit together with a Tisserand parameter with respect to Jupiter of 2.7 suggests that this body may in fact be a dynamically highly evolved comet, though no activity from it has yet been reported. The past extremely-low perihelion distance was previously noted by \cite{voknes12}, who computed that the surface of the asteroid could have reached 2500K. They assessed the effects of the denser solar wind, circumsolar dust and the Yarkovsky effect on the orbit of 2004~LG and found them to be negligible, so there is no obvious dynamical mechanism (unless the object does at times exhibit cometary outgassing) which would have prevented it reaching a very low perihelion distance.

There are no spectra of (467372) 2004~LG of which are aware. From NEOWISE, its diameter is 0.864 km ($H = 18.0$) and its albedo 0.146 \citep{maigrabau11}. The relatively high albedo makes it more likely to be an S-type than a C-type \citep{chamorzel75, tedwilmat89, masmaigra11}. S-types being associated with ordinary chondrites (e.g. \cite{brithobel92}) this could explain its survival, as a stony composition might be more resistant to a variety of destructive processes than a carbonaceous one. The NEO model of \cite{gramorjed18} gives a probability of zero of a JFC origin, with the 3:1 the most likely source (probability 0.33); thus (467372) is dynamically likely to be asteroidal.

To help determine if (467372) 2004 LG might indeed be a comet, we conducted an archival search for images of it using the Solar System Object Image Search (SSOIS) tool \citep{gwyhilkav12} at the Canadian Astronomy Data Centre (CADC) but without success. We also obtained five unfiltered 300 second exposures tracked at the expected sky motion of (467372) 2004 LG on 11 May 2019 with the 0.61~m T24 telescope on the itelescope.net network. The asteroid was approaching the Sun at a heliocentric distance of 1.11 AU at a phase angle of $62\degree$ with a JPL-predicted apparent magnitude of 19.9. The asteroid elevation was $45\degree$ and the Moon was below the horizon. The asteroid was visible in the stacked image at 1-sigma at an apparent magnitude of 20.4  but no cometary activity was detectable. Pan-STARRS images of this asteroid taken in June 2010 (Weryk, 2019, private communication) show a full-width at half-maximum (FWHM) perpendicular to the trail of 1.7'', nominally larger than that of the stars (1.46$ \pm 0.05$'') but it was not reported as cometary at the time, nor have any of the over 100 observations listed at the Minor Planet Center as of 27 May 2019 mentioned activity as far as we are aware.  At the moment it is unclear if (467372) 2004 LG is asteroidal or cometary but it presents a challenge to the SCD hypothesis.

\subsubsection{Phaethon}
\begin{figure}
\includegraphics[width=12cm]{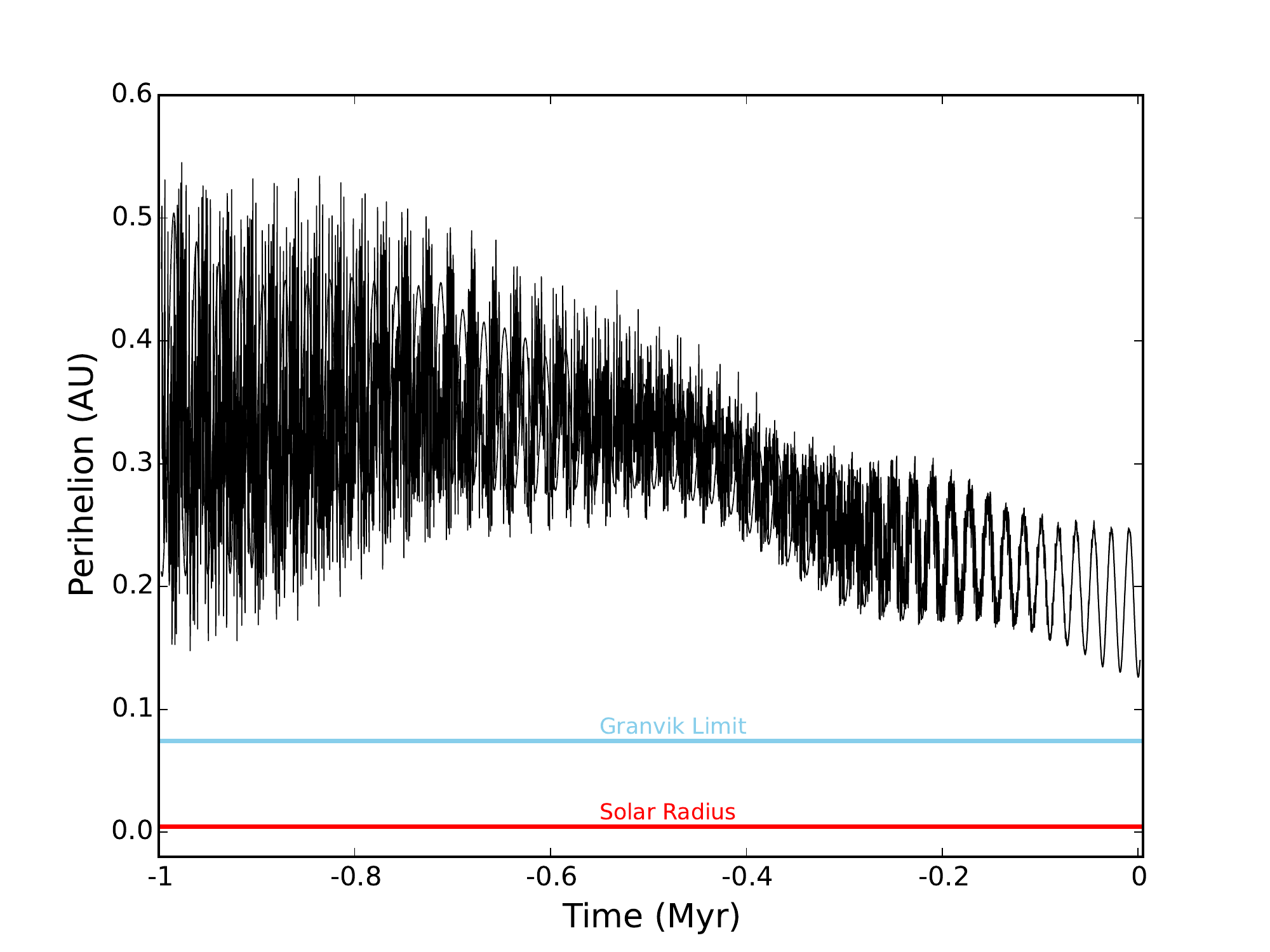}
\caption{Past evolution of the perihelion distance of (3200) Phaethon and 100 clones distributed within the orbital determination errors. The orbit is well-known and so all clones remain on very similar orbits. The perihelion distance fluctuates, remaining near the G16 limit but not crossing it. \label{fi:Phaethon} }  
\end{figure}

Asteroid (3200) Phaethon  ($a=1.27$~AU, $e=0.89$, $i=22.3\degree$) has a perihelion distance of 0.14~AU, low but outside the G16 limit. Unlike most asteroids, it is a known dust producer. It is recognized as the parent of the Geminid meteor shower \citep{whi83} and has been seen to produce dust (though not at cometary levels) when passing perihelion \citep{jewli10}. We could ask if Phaethon's dust production is due to dynamical changes in its perihelion distance: if it were at lower perihelion in the past, perhaps even inside the G16 limit for some time, this might provide a clue as to the nature of the supercatastrophic disruption process. Simulations of the orbit of Phaethon during the past million years show that its orbit has remained near but just above the 0.16~AU G16 limit (see Figure~\ref{fi:Phaethon}). Thus Phaethon may represent a boundary case, where its perihelion hovers above the supercatastrophic disruption limit for long times without crossing it. The dust production from Phaethon may then be the result of supercatastrophic disruption in slow motion. 

From the aforementioned simulations, we conclude that that past dynamical history of the near-Sun asteroids is broadly consistent with a removal/destruction process acting at small perihelion corresponding to the G16 limit.

\subsection{Near-Sun comets} \label{nearsuncomets}
If any of the current population of periodic low-perihelion comets is in fact an asteroid in the process of SCD, then it should be on an orbit consistent with an asteroid belt origin, that is, a short-period or Jupiter family orbit rather than Halley-type. The JPL comet list\footnote{ssd.jpl.nasa.gov/dat/ELEMENTS.COMET, retrieved on 23 March 2019}, there are only 10 short-period($<20$ yr) comets listed with $q < 0.3$ AU, and they are all SOHO-discovered comets with the exception of 96P/Machholz. Though the process of determining these orbits and of linking successive apparitions of SOHO comets is a difficult one, it is interesting that seven of these ten comets 1) are ungrouped 2) and have orbits with inclinations of less than $25\degree$ (Table~\ref{ta:nearsun}). So we have several good SOHO comet orbits consistent with an asteroidal or JFC origin, and thus with SCDing asteroids. Given that our sample of potentially SCDing asteroids consists almost exclusively of SOHO comets, we'll defer a more detailed discussion of them to the following section where the entire SOHO sample, not just those with the best-determined orbits, is examined.

\begin{table}[htbp]
    \centering
    \begin{tabular}{c|cccccc|ccccccc|l}
                   & \multicolumn{6}{c|}{Orbital elements}             & \multicolumn{7}{c|}{Source region probabilities } & \\
         name      & $q$  & $a$   & $e$   & $i$ & $\Omega$ & $\omega$ & $\nu_6$ & 5:2 & 2:1   & Hun   & 3:1   & Pho   & JFC   &  group \\
        \hline \hline
        321P       & 0.047 & 2.43 & 0.980 & 19.7 & 165     & 172      & 0.09 & 0.00 & 0.00 & 0.01 & 0.87 & 0.00 & 0.03 & Ungrouped \\
        322P       & 0.054 & 2.52 & 0.979 & 12.6 & 0       & 49       & 0.10 & 0.00 & 0.00 & 0.01 & 0.81 & 0.00 & 0.08 & Ungrouped (Kracht II)\\
        323P       & 0.048 & 2.61 & 0.982 & 6.5  & 322     & 355      & 0.04 & 0.00 & 0.02 & 0.00 & 0.25 & 0.00 & 0.70 & Ungrouped \\
        342P       & 0.053 & 3.04 & 0.983 & 13.3 & 43      & 59       & 0.00 & 0.00 & 0.04 & 0.00 & 0.09 & 0.00 & 0.86 & Ungrouped \\
        96P        & 0.124 & 3.03 & 0.959 & 58.5 & 94      & 15       & 0.02 & 0.32 & 0.33 & 0.00 & 0.20 & 0.02 & 0.11 & Machholz  \\
        P/1999 J6  & 0.049 & 3.10 & 0.984 & 26.6 & 82      & 22       & 0.00 & 0.00 & 0.08 & 0.00 & 0.12 & 0.00 & 0.79 & Machholz (Marsden)\\
        C/2002 R5  & 0.047 & 3.22 & 0.985 & 14.1 & 13      & 46       & 0.00 & 0.00 & 0.18 & 0.00 & 0.05 & 0.00 & 0.77 & Ungrouped (Kracht II) \\
        P/2002 S7  & 0.049 & 3.22 & 0.985 & 13.6 & 50      & 52       & 0.00 & 0.00 & 0.18 & 0.00 & 0.05 & 0.00 & 0.77 & Ungrouped \\
        P/2008 Y12 & 0.065 & 3.08 & 0.979 & 23.3 & 313     & 147      & 0.00 & 0.01 & 0.05 & 0.01 & 0.25 & 0.00 & 0.69 & Ungrouped \\
        C/2015 D1  & 0.028 & 4.94 & 0.994 & 69.6 & 96      & 235      &   -  &   -  &   -  &  -   &  -   &  -   &  -   & Ungrouped \\
       \end{tabular}
    \caption{The best determined near-Sun ($q<0.3$~AU) comet orbits from JPL.  Units of $q$ and $a$ are AU, angular elements are in degrees. The final column provides the near-Sun comet group. The seven columns just prior to that one give the probabilities of their source regions being the $\nu_6$, 5:2 or 2:1 resonances, the Hungarias, the 3:1 resonance, the Phocaeas or the Jupiter family of comets, according to the \cite{gramorjed18} NEO model. None of the comets have known $H$ magnitudes except for 96P (where $H\approx 15$) so these values are taken from the smallest size bin ($H\in[24.75 - 25.0]$). The orbital elements of C/2015 D1 (SOHO) are outside the region for which source probabilities are given by \cite{gramorjed18}. \cite{huiyekni15}'s orbit computation for C/2015 D1 differs somewhat from JPL's, and indicates it may in fact be a long-period comet. }
    \label{ta:nearsun}
\end{table}

\subsection{SOHO comets}
The Solar and Heliospheric Observatory (SOHO)'s Large Angle Spectrometric Coronagraph (LASCO) has been observing comets in the near-Sun region since 1996. Over 3000 comets, most passing within the G16 limit, have been observed and this sample may include SCDing asteroids. \cite{lamfaulle13} is the last comprehensive analysis of this sample, while \cite{jonknibat18} provide a useful overview of near-Sun comets: a brief review of the relevant orbital details follows. Below we will adopt the terminology proposed by \cite{kniwal13} where comets passing within the Sun's fluid Roche lobe are labelled 'sungrazers', while those passing further out are 'sunskirters'. Because the vast majority of these comets are only seen at a single perihelion passage, their semimajor axis and eccentricity are unknown in almost all cases, though their perihelion distance and inclination can be measured.

\subsubsection{Sungrazer comets - Kreutz group:} These have perihelia  around 2 solar radii, and because of their inclination $i$ and the large semimajor axes $a$ (e.g. $i \approx 144\degree$ $a\approx 64$~AU for the Great March Comet of 1843 - 1843 D1), they originate from the Oort cloud \citep{bielamstc02}. Most do not survive perihelion passage (though larger ones e.g. C/2011 W3 (Lovejoy) may (\cite{kre91} as cited in \cite{mar67}) and searches for them away from perihelion have been unsuccessful to date \citep{yehuikra14}.  Here we will take the sungrazer family to have originated from traditional cometary activity and/or fragmentation, because of its retrograde inclination and large semi-major axis.

\subsubsection{Sunskirters - Meyer group:} These comets have perihelia from 6 to 9 solar radii and high inclination ($i \approx 72 \degree$) \citep{sekcho05}. Many survive perihelion, at least briefly. \cite{sekcho05} consider it to be on a large aphelion orbit because arrivals are not clustered in time, suggesting a long orbital period, but neither its orbital period nor a potential parent body is known. We will consider it to be traditional cometary in nature, because of its apparent large aphelion and large inclination.

\subsubsection{Sunskirters - Marsden and Kracht groups (Machholz group):}  These comets have perihelia near 10 $R_\odot$ and inclinations of 10-35$\degree$ \citep{lamfaulle13}.  These comets typically survive their perihelion passage though no SOHO-discovered members have been seen by other telescopes. These comets are part of a group dynamically associated with 96P/Machholz~1 \citep{sekcho05, lamfaulle13} and which includes 2003 EH$_1$ \citep[an asteroid associated with the Quadrantids meteor shower;][]{jen04,wiebro05}, The Machholz complex is also associated with eight meteor showers at the Earth \citep[e.g.,][]{abewiejan18}.  The members of this group likely split from each other, but is this traditional cometary splitting or SCD? 

Sunskirting comets appear "stellar" in SOHO images.  \cite{lamfaulle13} report that only seven have short tails in their sample of 238. \cite{sekcho05} estimate that they are nearly inert objects that only survive one or two orbits, though brighter ones may survive longer. \cite{lamfaulle13} conclude that they are unlikely to be bare nuclei however, as the amount of reflected light corresponds to a cross-section of 10-100~km, unrealistically large compared to the estimated fragment size which is orders of magnitude smaller (meters to tens of meters). 

These comets could be an example of an SCD-produced family of fragments, despite its origin with a JFC: the NEO model of \cite{gramorjed18} includes a JFC source, and bodies i.e. comets originating from that source presumably need to undergo SCD as well, though this is not detailed in that reference. We note that 96P has only a 0.11 probability of originating from the JFC source in the \cite{gramorjed18} dynamical model (see Table~\ref{ta:nearsun}), despite being a JFC itself (though an atypical one because of its low $q$ and high $i$). Another point supporting the SCD origin of this group is the fact that it contains a large (inactive) asteroid (196256) 2003 EH$_1$ ($a = 3.12$~AU, $e=0.619$, $i=70.8\degree$, and which also has a \cite{gramorjed18} probability of a JFC origin of only 0.11) and the smaller members seen by SOHO might be small asteroids undergoing SCD. 

Arguably, the disintegration of 96P into the Machholz complex might be an expression of SCD, particularly since the primary members 96P and (196256) 2003~EH$_1$ are dynamically unlikely to have a JFC origin according to \cite{gramorjed18}. However, given the distinctly cometary nature of 96P, the largest member of this group (6.4 km diameter \citep{lamtotfer04}, corresponding to $H \approx 15$ for a typical low-albedo comet; versus 2-3~km for (196256) 2003~EH$_1$ depending on albedo: it has an absolute magnitude $H$ of $16.2$\footnote{https://ssd.jpl.nasa.gov/sbdb.cgi, retrieved 20 July 2019}), we conclude that this group is the result of traditional cometary fragmentation, though it may still inform the SCD process.

\subsubsection{ Ungrouped sungrazing and sunskirting comets:} \label{ungroupedSOHO}
There are 75 ungrouped SOHO comets reported by \cite{lamfaulle13}. With a wide variety of inclinations, and perihelia from 0.24 to nearly 40 solar radii, these are a mix of unrelated sungrazing and sunskirting comets.  Like other non-Kreutz comets, these are described as being meters to tens of meters in size, having  optical cross sections so high that they must be displaying coma of some sort, and many survive their perihelion passage \citep{lamfaulle13}. Though the spread in inclinations suggests an Oort cloud source, there is an excess of members at low inclinations. We will suggest here that the low-inclination population of ungrouped SOHO comets are in fact consistent with small asteroids undergoing SCD.

\begin{figure}
\begin{center}
\includegraphics[width=10cm]{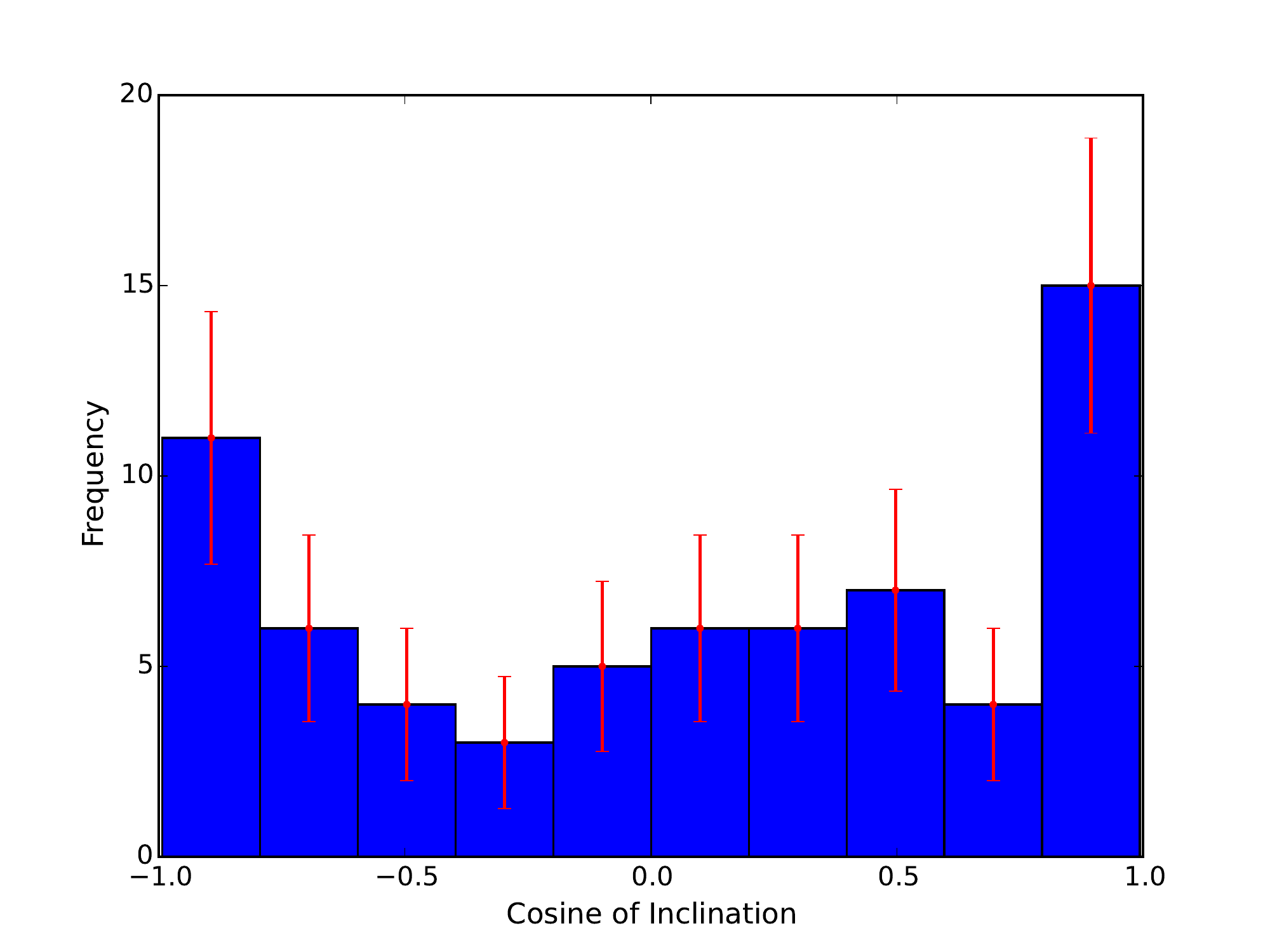}
\caption{Distribution of cosines of the inclinations of 67 ungrouped SOHO comets from \cite{lamfaulle13}. An isotropic distribution of orbit poles, such as one originating from the Oort cloud, would have a flat distribution. The observed distribution is consistent with a spherically symmetric component plus a component concentrated in the ecliptic plane. \label{fi:SOHO} }
\end{center}
\end{figure}

\cite{lamfaulle13} list 75 ungrouped SOHO comets recorded between 1996-2008.  There were 149 as of 2017 \citep{jonknibat18,batkni17} but few of the later ones have published orbits so we restrict our analysis here to the \cite{lamfaulle13} sample. Of the 75 with orbits, 12 were subsequently linked as multiple apparitions of comets 321P/SOHO, 322P/SOHO and 342P/SOHO and C/2002 R5 (SOHO), leaving 67 unique comets.  322P and 3 of these (C/2002 R5 (SOHO), C/2008 O6, C/2008 O7) are now considered a single member of the Kracht II group. 322P was the first SOHO comet to be observed by other instruments, and showed an absence of coma consistent with it being an asteroid \citep{knifitkel16}. The ungrouped comets do not show much clustering in arrival time, longitude of ascending node $\Omega$ or argument of perihelion $\omega$, and so are dynamically consistent with evolved asteroids.

Of our sample of 67, 38 of these have an inclination less than $90\degree$ and 29 more than $90\degree$.  A plot of the cosine of the inclinations is presented in Figure~\ref{fi:SOHO}. If the orbits were really randomly distributed on the sphere, as one might expect for Oort cloud comets, the distribution of $\cos i$ should be flat. But there is an excess near $\cos i$ of +1 and -1, indicating an excess of comets in the ecliptic plane. One explanation could be an observational bias towards comets at low inclinations. \cite{lamfaulle13} discuss observational bias in the LASCO observations but do not mention biases associated with the ecliptic plane. They do mention (as do \cite{batkni17}) that the short observational arc makes the orbit calculations difficult, with particular ambiguity in determining if the orbit is prograde versus retrograde. The plane of motion can be relatively well established but whether the comet passes in front of or behind the Sun is much harder to determine. So we will take the excess of ungrouped SOHO comets in the ecliptic plane to be real, and use them to calculate the rate at which they are produced for comparison with near-Earth asteroid models.

We can set an upper limit on the number of ungrouped comets coming from the asteroid belt, by assuming that the $\cos i$ distribution of ungrouped comets consist of a uniform background of Oort cloud comets plus an excess asteroidal component. For the sake of argument, we will take the excess of comets at inclinations below $25.8\degree$ and above $154.2\degree$ (the first and last bins of Fig~\ref{fi:SOHO}, since $\cos^{-1} 0.9 = 25.8\degree$) to be entirely  asteroids moving on prograde orbits. In this case, we have $\sim 15$ potentially SCDing asteroids over our sample time frame of $\sim 12$ years, assuming the SOHO comet detection efficiency is not too far from unity, or about one per year.

So if SOHO has been observing about one potentially SCDing asteroid per year, how does this rate compare with that expected? \cite{yegra19} calculate that a single asteroid of diameter greater than 0.5~km breaks up via SCD every 2000~yrs. Given a power-law cumulative number distribution for small NEAs proportional to $D^{-2.7}$ \citep{brosparev02} with $D$ the asteroid diameter, that extrapolates to one 30-meter diameter asteroid per year, consistent with SOHO comet size estimates of 'meters to a few dozen meters' \citep{lamfaulle13}.

Only a few ungrouped comets have sufficiently good orbits to use \cite{gramorjed18}'s model to determine the probability of their originating in a particular source region: these are given in Table~\ref{ta:nearsun}. Though the Jupiter-family comets is the most likely source for many of them, all show appreciable (0.2 - 0.8) probabilities of a main-belt origin as well.

We conclude that the ungrouped sunskirting comets are consistent with the SCD process both in terms of their rate and their orbits. Since some larger ungrouped SOHO comets survive at least one perihelion passage (e.g. 322P, 150-320m diameter), while smaller ones ($\sim 10$m) typically do not (See Table 4 in \cite{lamfaulle13}), this suggests a boundary between $\sim$10 and $\sim100$ m sizes, the smaller ones unable to survive even a single perihelion, implying $\sim 10$~m of material is lost per perihelion. 

\subsection{Meter-sized meteoroids}
\label{methods-meter}

Meter-sized asteroids on near-Sun orbits could either be small objects created (presumably by collisions) in the asteroid belt and driven to small $q$ by the same processes as large asteroids, or they could be debris from the SCD or other processes.

\begin{figure}
\includegraphics[width=8cm]{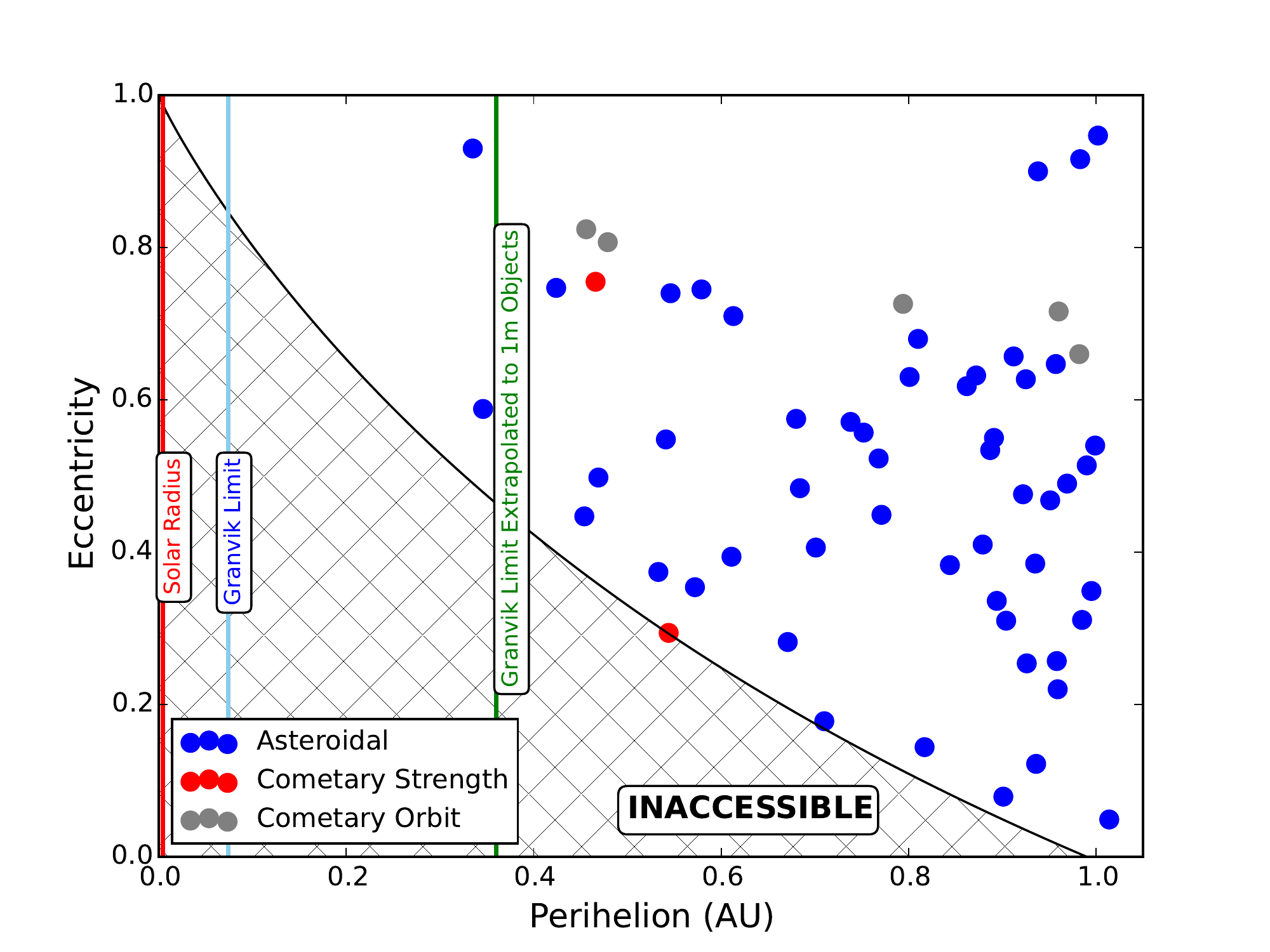}
\includegraphics[width=8cm]{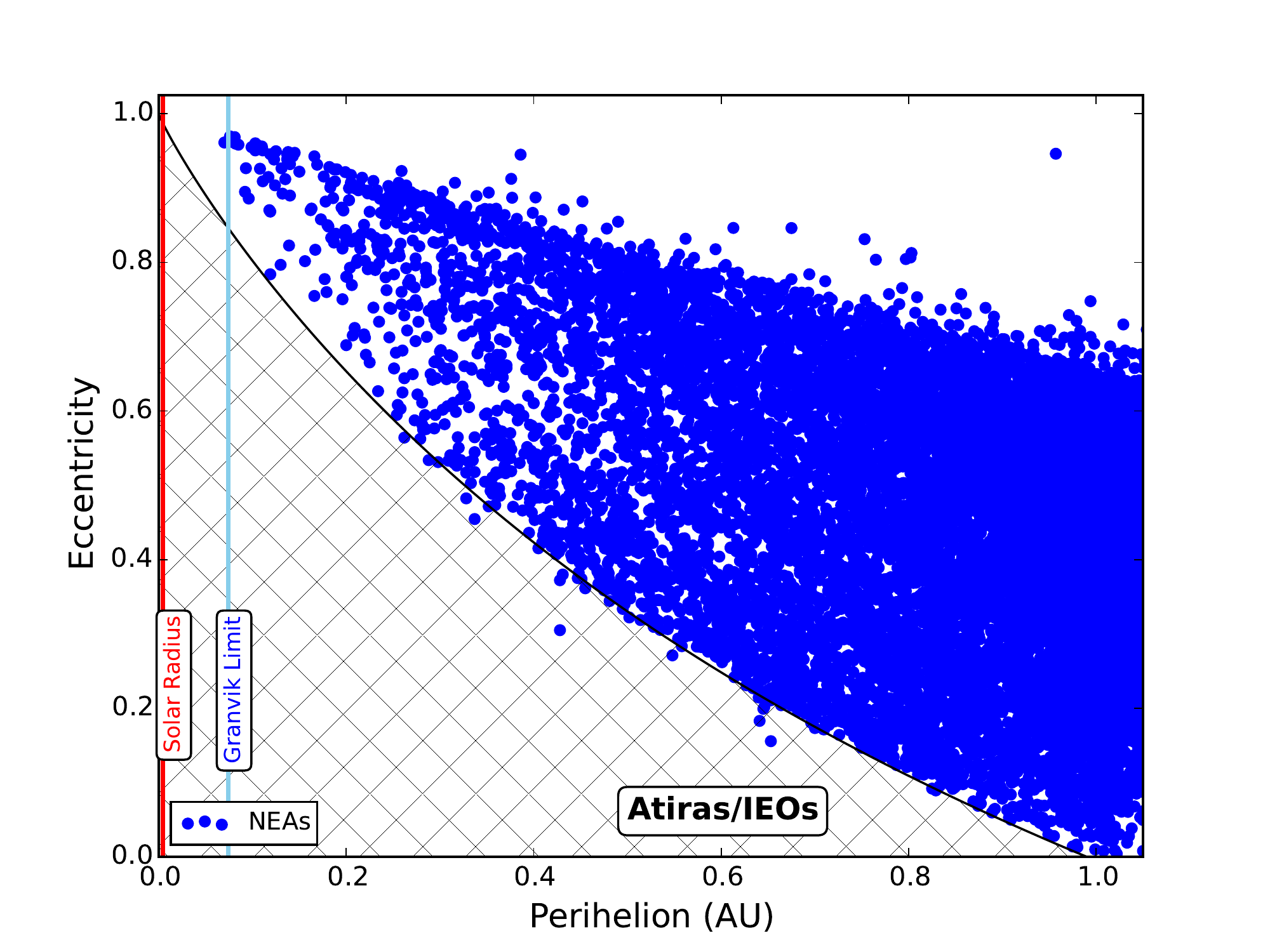}
\caption{Perihelion versus eccentricity for the 59 meter-class fireballs of \cite{browiecla16} (left panel) and the known near-Earth asteroids (right panel).  The shaded area indicates orbits which are entirely inside the Earth's (Interior-Earth Objects (IEOs) or Atiras). The G16 limit is indicated on both panels. The left-hand panel also includes an extrapolation of the G16 to meter sizes, which corresponds closely to the lowest perihelion fireballs so far observed. \label{fi:metersized} }
\end{figure}

Here we will compare the sample of meter-class impactors observed at Earth to the near-Earth asteroid population near the Sun. The largest well-characterized sample of meter-sized orbits is that of \cite{browiecla16}. They analyzed the orbits and the in-atmosphere characteristics of 59 fireballs produced by meteoroids with pre-atmospheric diameters of 1 m or larger. This sample's distribution of perihelion $q$ versus eccentricity $e$ is shown in Figure~\ref{fi:metersized}, where a deficit of bodies in the near-Sun region can be seen. The fireballs produced by impactors of this size are bright and often seen during daylight, and so radiants located in the near-Sun region are not strongly biased against. 

Indeed, the majority of this sample comes from US Government sensor measurements which do not show any significant day-night asymmetry\footnote{https://cneos.jpl.nasa.gov/fireballs/, retrieved 03 Oct 2019}.   The sample also includes fireballs collected by different ground-based camera networks  so it is not a completely uniform sample, but is not expected to be strongly biased against near-Sun radiants or daytime meteors. The sample is comprised mostly of asteroidal material with only a few meter-class fireballs with the characteristics of cometary material, but none were observed to have perihelia particularly near the Sun. The accuracy of individual orbital measurements from US Government sensor data has been shown to have wide variability \citep{Devillepoix2019}, but the perihelion distance tends to be among the more robust orbital elements with respect to meteor trajectory errors.

For comparison, the known near-Earth asteroid populations is plotted in the right-hand panel of Figure~\ref{fi:metersized}. The near-Earth asteroid catalog was downloaded from NeoDys\footnote{
http://newton.dm.unipi.it/$\sim$neodys2/neodys.ctc, retrieved 06 Jun 2018} and contains 17786 NEAs. We exclude 2015~KP$_{157}$ which has a 2 day arc with only 11 observations: its nominal $q$ is $0.053$~AU but with large uncertainty.  The minimum perihelion distance remaining among the NEAs is then 2005~HC$_4$, with $q = 0.07$~AU . 

On both panels of Figure~\ref{fi:metersized}, the region interior to the Earth's orbit has been indicated with cross-hatching. This region is strictly unobservable in the case of fireballs, which clearly must cross Earth's orbit; for the near-Earth asteroids;  this region is not completely inaccessible but difficult to sample telescopically.

The near-Earth asteroid sample has not been debiased in Figure~\ref{fi:metersized}, and so the two are not strictly comparable. A proper assessment of the observational effects biasing both samples would be required for a complete analysis, but we do not attempt that here. However it is clear that the data in hand is broadly consistent with an absence of meter-sized bodies in the near-Sun region. If we extrapolate the G16 Figure 2 size dependence to meter sizes, we find that a 1~m object wouldn't survive inside 0.36~AU. This distance is indicated on Figure~\ref{fi:metersized} and corresponds closely with the smallest $q$ fireballs observed. This is an interesting clue, and may indicate that the supercatastrophic process extends to even smaller sizes than G16 originally proposed, though it may simply be due to small number effects as well.

Unfortunately there is no additional compositional information for most of these fireballs. The two with the lowest perihelion for which meteorites were recovered are Maribo ($q=0.479$~AU, a CM2 chondrite \citep{spuborhaa13}) and Sutter’s Mill ($q=0.456$~AU, a CM regolith breccia \citep{jenfriyin12}). These compositions are puzzling at first glance. Carbonaceous chondrites being weaker than the more abundant ordinary chondrites \citep{cotaspgar16}, they should be more vulnerable to most disruption processes. G16 also found that high albedo asteroids could survive closer to the Sun than those with low albedos, but carbonaceous meteorites are associated with lower albedo asteroids \citep{chamorzel75, tedwilmat89, masmaigra11}. In contrast to this, the bulk density of CM meteorites ($2.2$~g~cm$^{-3}$ \cite{briyeohou02}) is lower than that of ordinary chondrites (3.6-3.9~g~cm$^{-3}$),  achondrites (3.2~g~cm$^{-3}$), and of other carbonaceous chondrites except CI (2.1~g~cm$^{-3}$), which could allow them to better survive thermally-driven destruction processes.

However, the fact that our two lowest-$q$ meteorites are CMs is likely simply a result of poor statistics. 1) The three other carbonaceous meteorites with known orbits do not have low perihelia. Orgueil had $q=0.87 \pm 0.01$~AU \citep{gouspubla06}, Murchison had $q=0.99 \pm 0.01$ AU \citep{halmci90},  and Tagish Lake had $q=0.891\pm0.009$~AU \citep{brohilzol00}. 2) The cosmic ray exposure ages of Maribo ($0.8-1.4$ Myr \citep{haagrabis12}) and Sutter's Mill ($0.082 \pm 0.008$ Myr \citep{niscafham14}) are very young compared to carbonaceous chondrites over all (millions to tens of millions of years, \cite{schsch00}) so these may well have survived simply due to their youth. 3) Sutter's Mill and Maribo were both particularly large impactors. Maribo (3.3m diameter) had the fourth largest initial size of any of the 59 meteorite-producing fireballs with well-measured orbits in \cite{browiecla16},  while Maribo (1.1 m) was in the top 1/3. These are both atypically large precursor meteoroids which may help to explain why they could survive to such relatively small $q$. 

Since meter-sized bodies are driven into the near-Sun region by essentially the same dynamical effects as those which drive km-sized near-Earth asteroids there, we conclude that meter-sized bodies supercatastrophically disrupt by processes ostensibly similar to those affecting km-sized ones.  This also suggests that km-sized bodies do not break up into m-sized bodies, or at least only do so temporarily before those pieces are themselves destroyed/removed. This has implications for the underlying mechanism as some possibilities, such as tidal disruption, spin-up, the explosive release of volatiles or collision with another asteroid, would all produce large fragments which are not seen. Processes which produce small fragments are thereby favoured, such as thermal cracking or meteoroid bombardment.

\subsection{Millimeter-sized meteors}
\label{methods-millimeter}

The sample of millimeter-sized particles coming from the near-Sun region provides information similar to the meter-sized sample, but with some additional complications.

Firstly, there are a number of known comets with perihelion in the near-Sun region, and thus contaminate the sample of SCD-produced asteroid meteoroids (if any). On top of this are almost certainly long since extinct and/or disrupted comets, whose dust may remain though the parent objects are now gone. This was less of an issue for meter-sized meteoroids.  Comets are not important sources of meter-class meteoroids: meteor showers are not expected to contain meter-sized objects \citep{Beech1999}, searches for meter-sized fragments of Phaethon (Geminids shower parent have been unsuccessful \citep{jewasmyan19,tabwieye19,yewiehui19}, nor are any such meter-sized shower members observed associated with showers \citep{browiecla16}, with the exception of the Taurid meteor stream \citep{Spurny2017}. The cometary ejection processes driven by water sublimation are limited to lifting particles smaller than $\sim$10~cm from the nucleus surface \citep{whi51,whihue76, Beech1999}. But millimeter-sized cometary meteors with low perihelia are common and must be carefully addressed.

Secondly, millimeter-sized meteoroids are subject to significant Poynting-Robertson (PR) drag. This is a relativistic effect due to the asymmetric re-radiation of sunlight, and causes these particles to spiral into the Sun and systematically reduces their perihelion distances: such dynamical effects must also be taken into account.

Our sample of millimeter-sized near-Sun meteors is collected with the Canadian Meteor Orbit Radar (CMOR) \citep{webbrojon04, Jones2005, Brown2008}. Located near London, Canada, CMOR is a backscatter radar comprising a 15 kW transmitter operating at 29.85 MHz and five separate receiver stations. Meteor echoes detected at the main site plus two or more stations produce velocity measurements of approximately 5000 meteor echoes daily. Meteoroid orbits measured by CMOR have size limits which are strongly velocity dependent. The lower size limit for detection at 70 km/s is sub-mm to almost 5 mm at 12 km/s \citep{Ye2016}.

\begin{figure}
\includegraphics[width=8cm]{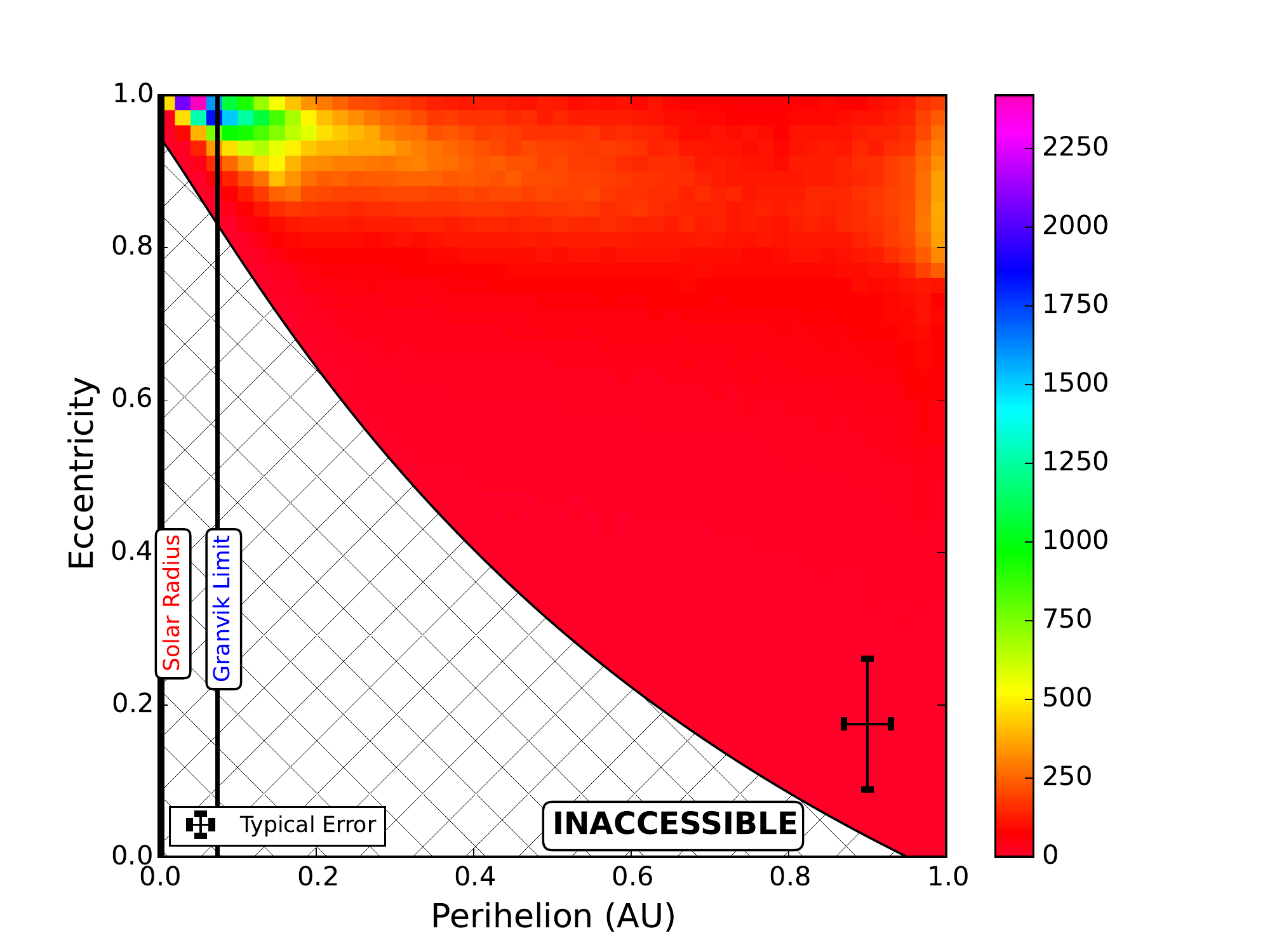}
\includegraphics[width=8cm]{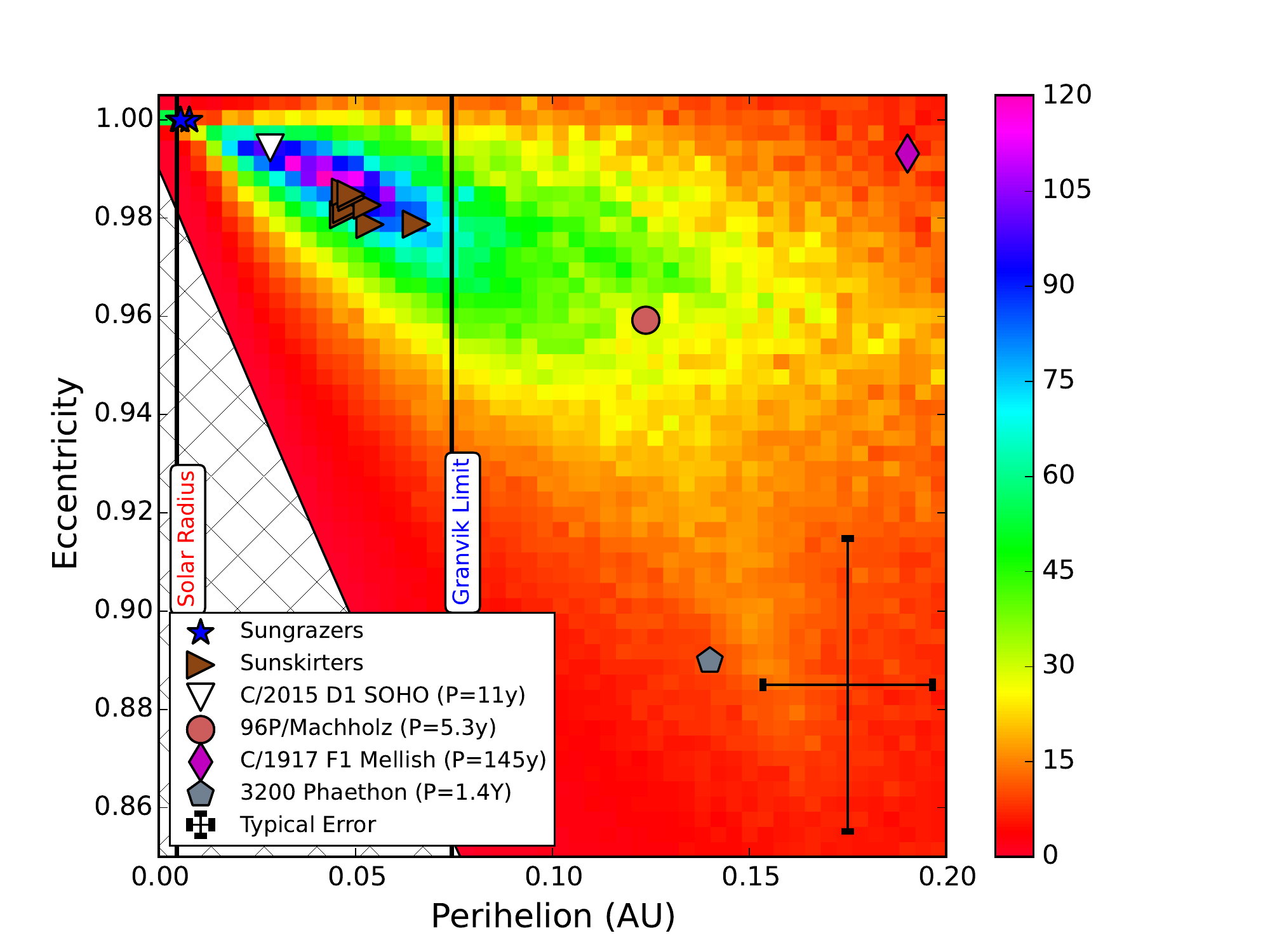}
\caption{Debiased number of CMOR collected meteors, for all meteors with $q<1$AU (left panel) and for the near-Sun region (right panel). The G16 limit and the radius of the Sun are indicated by vertical lines. The median error bars for an individual meteor are shown at the lower right. The distribution of comets with $e < 1$ and period $P < 1000$~y is superimposed for comparison. There is clearly a near-Sun meteor population but much of it is cometary in origin. The hatched region cannot be sampled from Earth because these orbits do not intersect our planet. \label{fi:millimetersized} }
\end{figure}

The advantage of a meteor radar over optical camera meteor detection is that the radar --being able to operate with equal efficiency both day and night -- is not biased against the near-Sun region. Thus our sample includes low-perihelion meteors collected both inbound to and outbound from the Sun.  Figure~\ref{fi:millimetersized} shows a density plot of the meteors collected during 2011-2019. The sample of meteors with $q < 1$~AU contains 2,027,678 meteors. These represent the highest-quality meteors in the sample. They were  selected because they were detected by a minimum of four of the six CMOR stations, had a consistent measured inflection pick at all stations compared to the best fit time of flight velocity solution and had in-atmosphere time of flight speed within the pre-t0 speed and its uncertainties, based on the hybrid KDE method described in \cite{mazur2019}. The pre-t0 speed is an independent single-station estimate of speed for specular echoes which can be used to check the time of flight method. Hence, when the time of flight speed and pre-t0 speed agree to within uncertainty we have high confidence in the measured speed (and trajectory). 

Meteor radars measure the number of meteors passing through a particular atmospheric area; to make this comparable to the telescopic sample of asteroids, which is a volume sample, each meteor is assigned a debiasing weight inversely proportional to the collision probability of its orbit with our planet\citep{opi51}. An additional term corrects for the relative geometry of the meteor radiant and the gain pattern of the radar. This term is inversely proportional to the instantaneous effective collecting area of the radar as described in \cite{Kaiser1960, Brown1995}. We will refer to the meteoroid orbit sample for which these corrections have been made as the 'weighted' or 'debiased' sample, and it is suitable for direct comparison with the usual telescopic sample of asteroids, if no size threshold is considered.

Figure~\ref{fi:millimetersized} shows the distribution of meteoroid orbits in the inner solar system as measured by CMOR. Of the most interest to use here is the concentration of meteoroids in the upper left corner. There is undoubtedly a concentration of meteor orbits with low perihelia and we can immediately conclude that millimeter-sized meteoroids can survive in the near-Sun region even if larger asteroids may not. But many of these meteors are cometary in origin, and thus not the result of asteroidal SCD. 

One can first search for signs of the SCD process by looking for near-Sun meteor showers with no known parent body. This was attempted by \cite{yegra19} who found that a disrupted 0.5 km asteroid could supply such a stream. However,they found more meteoroid streams than was compatible with the expected rate of asteroid SCD. Here instead we will look at the broad non-localized distribution of meteoroids consistent with the disruption of meter-class asteroids, which follow essentially the same dynamical pathways as km-sized asteroids into the near-Sun region. This is motivated by the ungrouped SOHO comets discussed in section~\ref{ungroupedSOHO}, and the fact that the debris component (if any) of small asteroids is less subject to the vagaries of small numbers than that of km-class asteroids.

Radar meteor measurements do not provide sufficient information to distinguish asteroidal from cometary meteors based on physical ablation behaviour. So we are reduced to trying to disentangle cometary dust from asteroidal dust on the basis of their orbits. We will assess the existence of an 'asteroidal SCD' component near the Sun by examining a restricted region of orbital space calculated to exclude 1) those regions occupied by known periodic comets over the last 10,000 years and 2) the strong near-Sun meteor showers (which are from comets).  Once we discuss how we obtain this sample, we will return to a discussion of the origin and properties of any dust found there.

\subsubsection{Determining an uncontaminated region} \label{uncontaminated}

\begin{figure}
 \begin{center}
 \includegraphics[width=20cm]{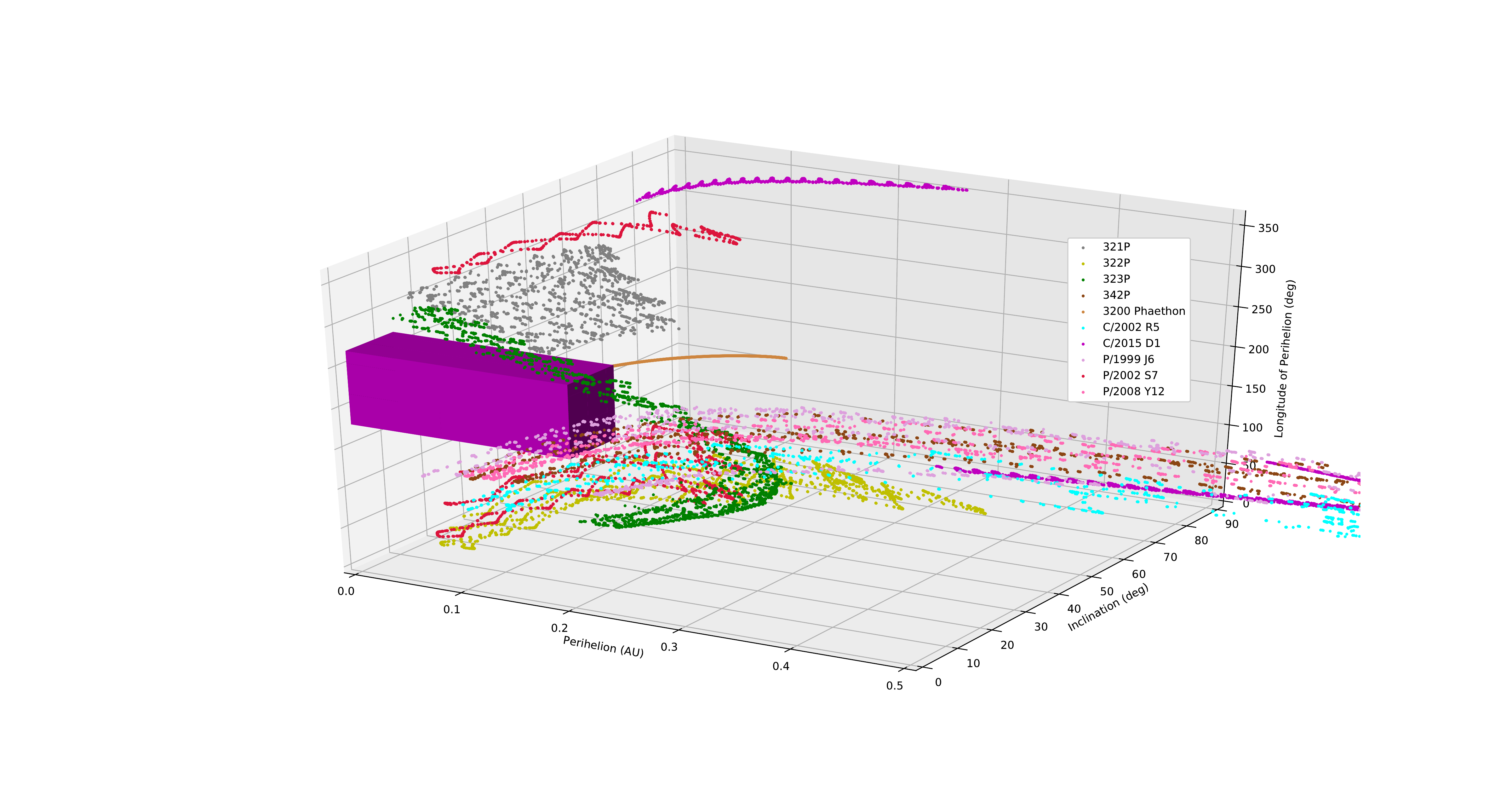}
\caption{The past evolution of the perihelion distance $q$, inclination $i$ and longitude of perihelion $\varpi$ for our selected comets (see Table~\ref{ta:nearsun}) over the last 10,000 years. Known comets move through a substantial fraction of near-Sun space during this time frame, complicating our attempts to find a region where cometary dust doesn't dominate. Our choice of 'uncontaminated region' is shown in purple. \label{fi:cometsims}}
\end{center}
\end{figure}

Our first step examines contamination by other comets of the near-Sun region  by integrating known periodic comets backwards in time for 10,000 years, and then excluding any regions of phase space that they visited from our sample. This is a conservative filter, since meteoroids shed by comets evolve dynamically in much the same way as their parents, and a meteoroid ejected in the past does not remain on the orbit it was deposited on, but rather follows the parent comet roughly through phase space. Here we consider the 10 known short-period comets with $q < 0.3$~AU, discussed earlier in Section~\ref{nearsuncomets} (see also Table~\ref{ta:nearsun}). We integrate the nominal orbits with the RADAU15 \cite{eve85} algorithm with a tolerance of $10^{-12}$. The influence of the Sun and eight planets is included as are the effects of General Relativity (post-Newtonian approximation). Cometary non-gravitational forces are are not considered: they are not well-known for these comets, and though possibly strong during perihelion passage, they are short-lived and have a very short lever-arm as well. Our purpose here is not to perform an exhaustive simulation of the near-Sun environment, only to avoid the most obvious sources of dust contamination. Figure~\ref{fi:cometsims} presents the time evolution of these comet orbits.

For our second step, we exclude the two strongest near-Sun meteor showers, the Geminids ($q=0.137$~AU, $i=23.2\degree$) associated with asteroid (3200) Phaethon, and the Machholz complex showers, which includes the Daytime Arietids ($q =0.074$~AU, $i=30.6\degree$, $\varpi=106.7\degree$), north ($q=0.096$~AU, $i=24.8\degree$, $\varpi=108.9\degree$) and south ($q=0.058$~AU, $i=31.5\degree$, $\varpi=100.1\degree$) $\delta$ Aquariids. The orbital elements above are the CMOR values for these showers reported in \cite{browonwer10}.

\begin{figure}
 \begin{center}
\includegraphics[width=8cm]{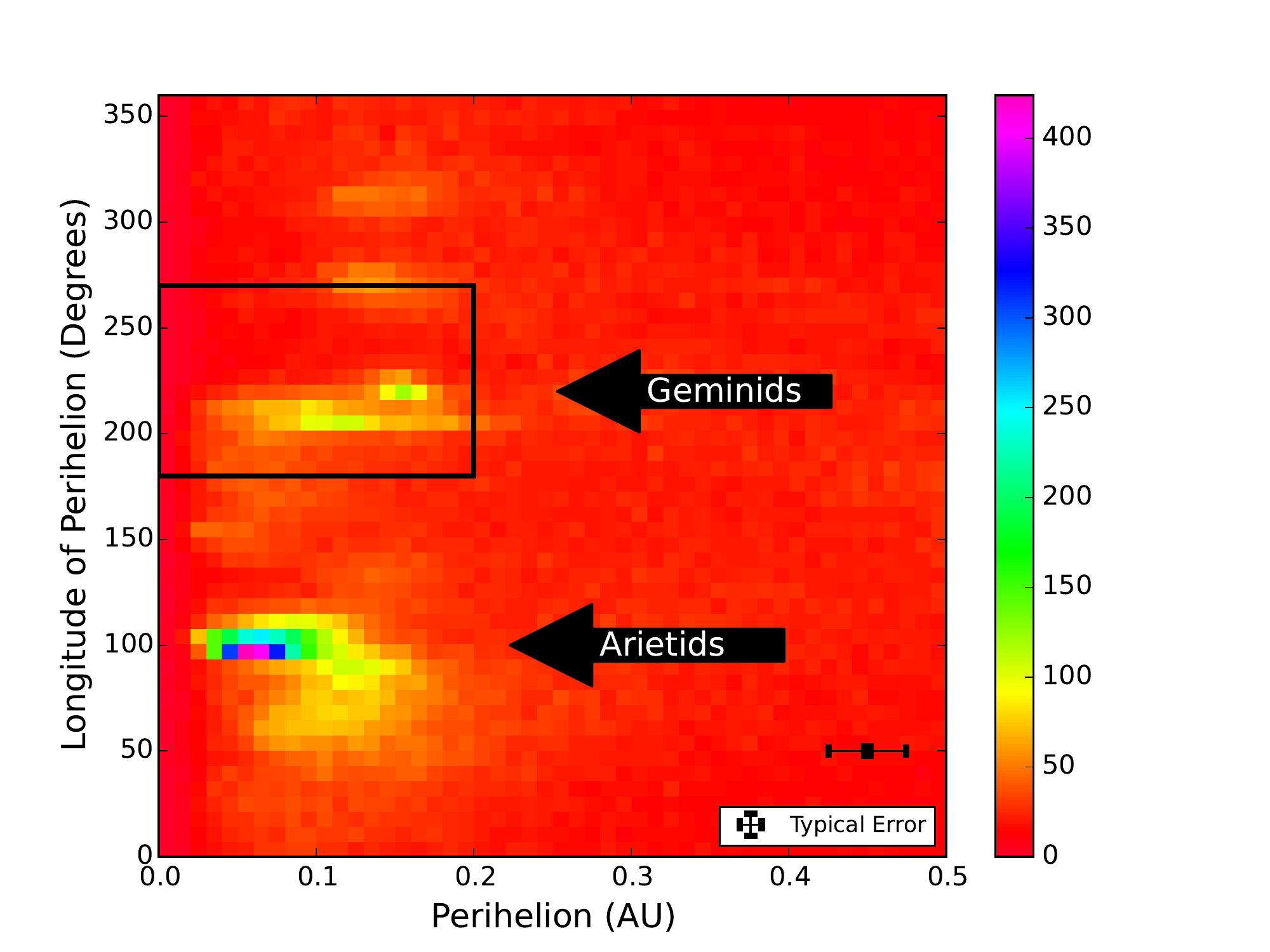}
\includegraphics[width=8cm]{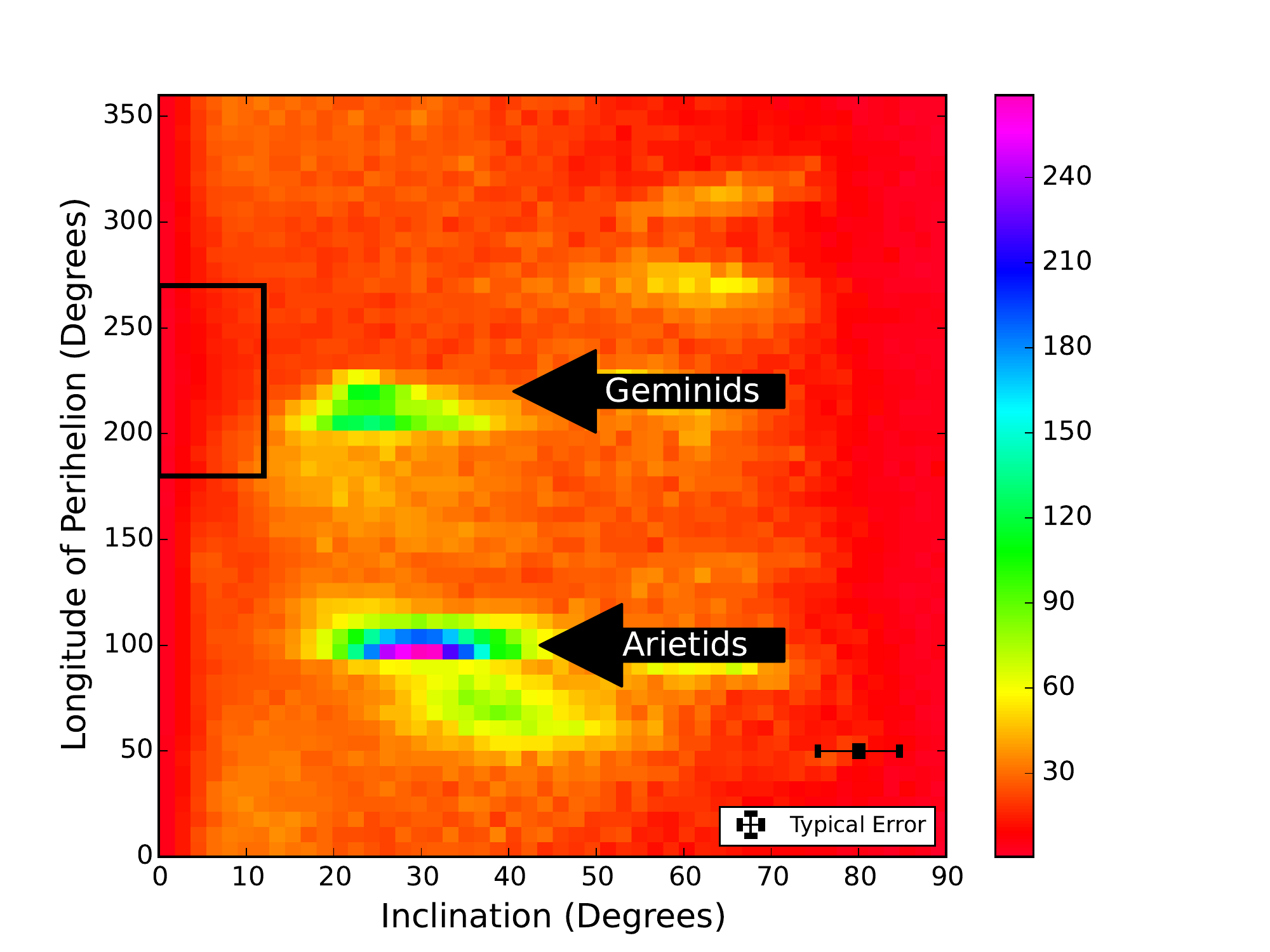}
\caption{Debiased meteor density plot for the near-Sun region for all perihelion less than 0.5~AU and inclinations less than $90\degree$. The arrows indicate the location of two major meteor showers, the Geminids and daytime Arietids. The region of our sample, selected both to avoid contamination by known meteor showers and to avoid potential contamination by known low-perihelion comets, is indicated by the black rectangle. \label{fi:meteors-q-lper}}
\end{center}
\end{figure}

A plot of the density of CMOR meteor orbits as a function of perihelion distance $q$ and longitude of perihelion $\varpi = \Omega + \omega$ is shown in Fig.~\ref{fi:meteors-q-lper}. Longitude of perihelion $\varpi$ is chosen as it is more closely conserved over time than longitude of the ascending node.

Much of the near-Sun region is visited by comets in our sample (Fig~\ref{fi:cometsims}), and there are limited choices of uncontaminated regions. Near $\varpi \sim 100\deg$ we see the Machholz complex comets associated with the daytime Arietids meteor shower, and the presence of this strong shower, along with associated contaminating comets (Fig.~\ref{fi:cometsims}) will lead us to exclude the region with $0 < \varpi < 180\degree$. The Geminids are visible near $200\degree < \varpi < 230\degree$, while the Daytime Arietids and other Machholz complex meteors are at $50 \degree < \varpi < 120\deg$. We can see from Figure~\ref{fi:cometsims} that regions with $270\degree < \varpi < 360\degree$ may have low inclination contamination from 321P ($i = 20\degree$) and 323P ($i = 6\degree$), two small ungrouped SOHO comets. These are unlikely to contribute much dust because of their size, and indeed there seems to be little dust apparent in this region of Fig.~\ref{fi:meteors-q-lper}. We will see later that these may in fact be SCDing asteroids (see section~\ref{ungroupedSOHO}), but for safety we will exclude the regions they visit. The region of $180\degree < \varpi < 270\degree$ is free of low inclination comets, but does contain Phaethon and the Geminid shower ($q=0.137$~AU, $i=23.2\degree$, \cite{browonwer10}). The compactness of the shower in inclination and the motion of Phaethon's dynamics suggests we can remove Geminid contamination by excluding inclinations above 12 degrees. Though the Geminids do not appear to create much dust at $\varpi > 230 \degree$, to avoid creating an awkwardly-shaped sample region,  we will exclude $i > 12\degree$ across our sample.

\subsubsection{Final sample region}

\begin{figure}
 \begin{center}
\includegraphics[width=8cm]{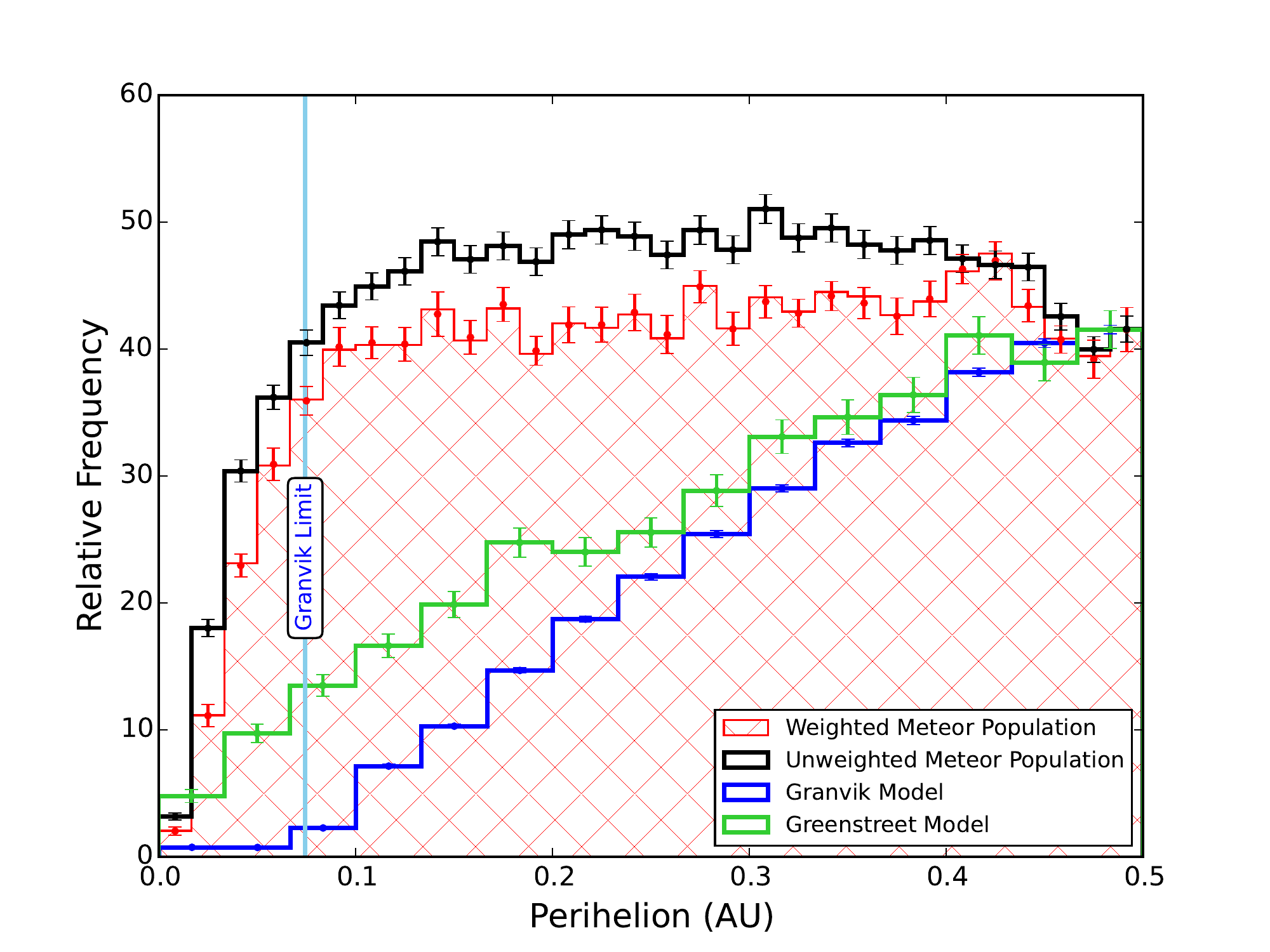}
\includegraphics[width=8cm]{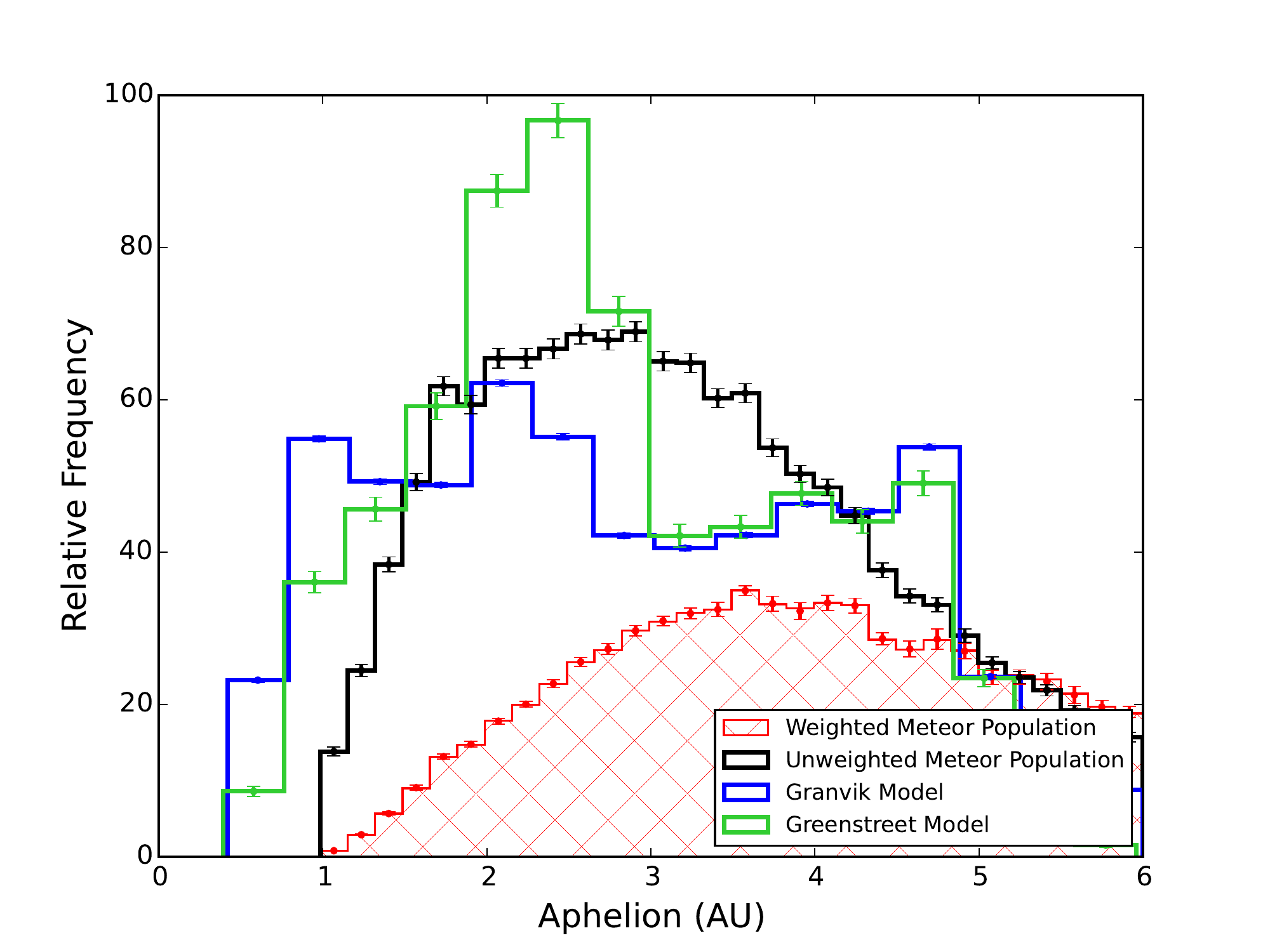}
\caption{Perihelion and aphelion distribution for our sample of near-Sun radar meteors, The \cite{grengogla12} and \cite{gramorjed18} models are included for comparison. There is an overabundance of
  near-Sun meteors at low perihelia compared to NEAs, including considerable material inside the G16 limit. All models are normalized to the same value at $q=0.5$~AU for perihelion distance, and at $Q=5.2$ for aphelion distance. \label{fi:meteors}}
\end{center}
\end{figure}

The aforementioned considerations lead us to select our notionally uncontaminated region as $0 \degree < i < 12\degree$, $180\degree < \varpi < 270\degree$ shown in purple in Figure~\ref{fi:cometsims}. A histogram of the number of meteor orbits in this region is shown in Figure~\ref{fi:meteors}. There is definitely a population of millimeter-sized meteoroids in this region of near-Sun space.

Figure~\ref{fi:meteors} tells us that millimeter-sized asteroidal meteoroids are abundant in the inner Solar System. They become more abundant relative to NEAs as $q$ decreases, and they can survive well within the G16 limit. These properties all support the hypothesis that SCDing asteroids disrupt into millimeter-sized fragments, fragments which are themselves relatively resistant to the disruption process.

Extracting additional information from these orbits is difficult as 1) despite our debiasing (Section~\ref{methods-millimeter}) the need for collision with the Earth imparts a complex geometrical constraint on the data, and 2) these orbits are highly-evolved. We do however note that the material in our subsample is on orbits with aphelia $Q$ near 4~AU (see Fig~\ref{fi:meteors} right panel), consistent with evolved material, either from JFCs or the main belt, that is moving inwards under PR drag.

Our model of cometary contamination did not include 2P/Encke as its perihelion distance ($q=0.33$~AU) is nominally outside the limit of 0.3~AU chosen in Section~\ref{uncontaminated}, but could it be a contributor? Its aphelion distance $Q$ is near 4 AU (where our meteor sample peaks, Fig~\ref{fi:meteors}), and it is associated with the Taurid meteor shower \citep{whi40}. 2P/Encke's inclination ($i=11.8\degree$) and longitude of perihelion ($\varpi = 161.1\degree$) are near but outside our sample region, and Fig.~\ref{fi:meteors-q-lper} shows little meteor activity near the location in question.  We conclude that it is not a significant contributor to our sample. However it does illustrate the difficulty in determining the origin of this material.  Though we cannot categorically exclude extinct/disrupted JFCs as the source of this dust, it cannot obviously be linked to known cometary parents and is consistent with the resonant processes that drive NEAs out of the main belt and into the Sun \citep{farfrofro94, glamigmor97}.

\begin{figure}
 \begin{center}
\includegraphics[width=14cm]{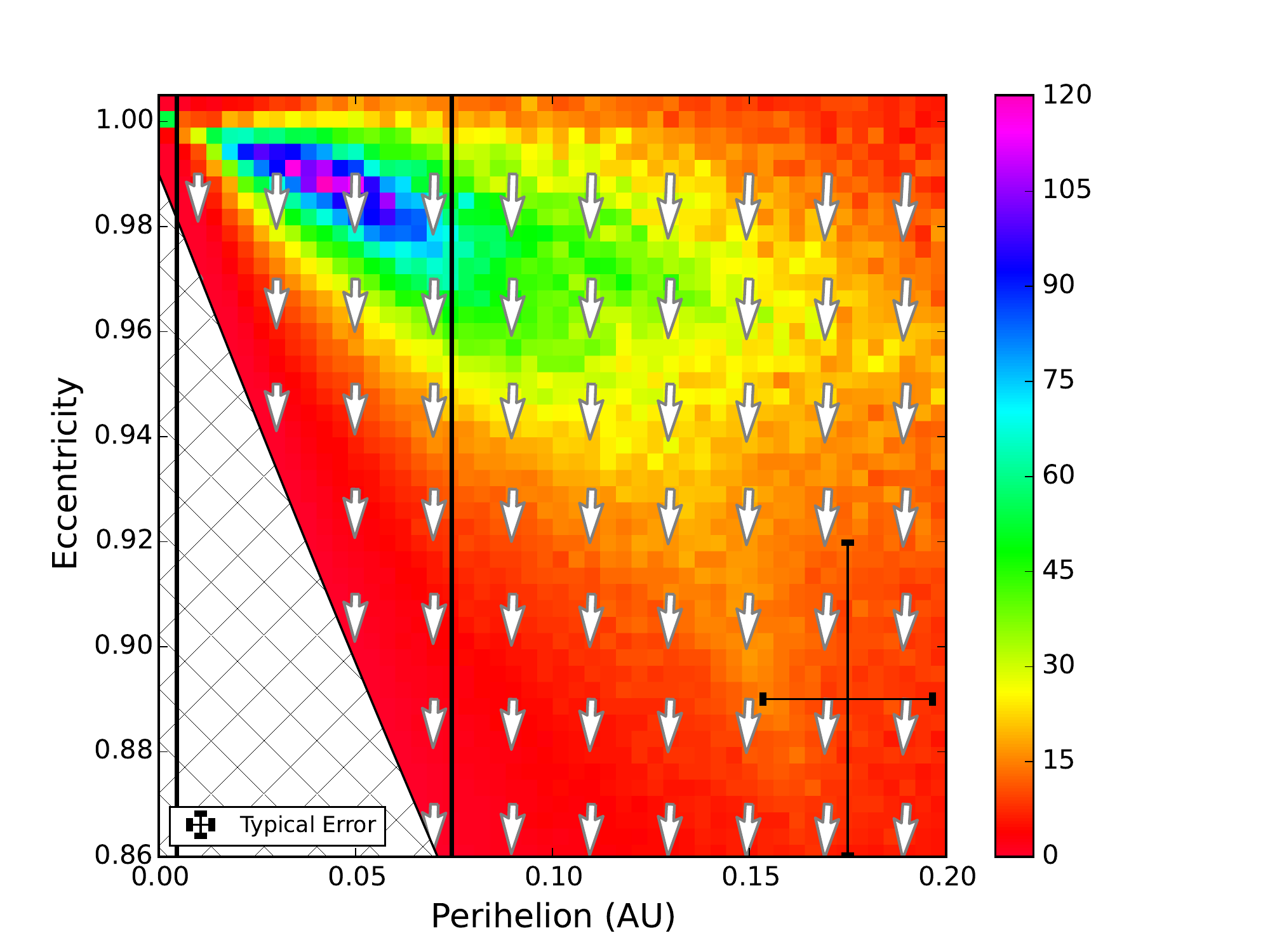}
\caption{Near-Sun meteor distribution with arrows indicating the direction of evolution under PR drag. The length of the arrows is proportional to the log$_{10}$ of the magnitude of PR drag. The direction of the arrows indicates that PR drag is removing material from our sample region, by circularizing their orbits faster than it decreases their perihelion distance.  Median measurement errors for meteors in this sample are represented by the black cross in the bottom right part of the figure.\label{fi:PRdrag}}
\end{center}
\end{figure}

It is worth noting that these meteoroids cannot have been put in place solely through the effects of Poynting-Robertson (PR) drag on meteoroids produced elsewhere in the Solar System. This is because PR drag, though it does cause meteoroids at these sizes to spiral in towards the Sun, also causes the eccentricity $e$ to monotonically decrease, and very rapidly relative to the rate of decrease of $q$ in our region of interest. \cite{wyawhi50} provide analytic expressions for PR drag evolution and show that particles follow
\begin{equation}
    q = \frac{C e^{4/5}}{1+e}
\end{equation}
independent of $a$, where $C = q_0 e_0^{-4/5}(1+e_0)$ and $q_0$ and $e_0$ are the initial values of the perihelion and eccentricity. These produce very steep curves in our region of interest as shown in Figure~\ref{fi:PRdrag}. As a result, PR drag is actively removing material from our sample rather than injecting it.

\begin{figure}
 \begin{center}
\includegraphics[width=14cm]{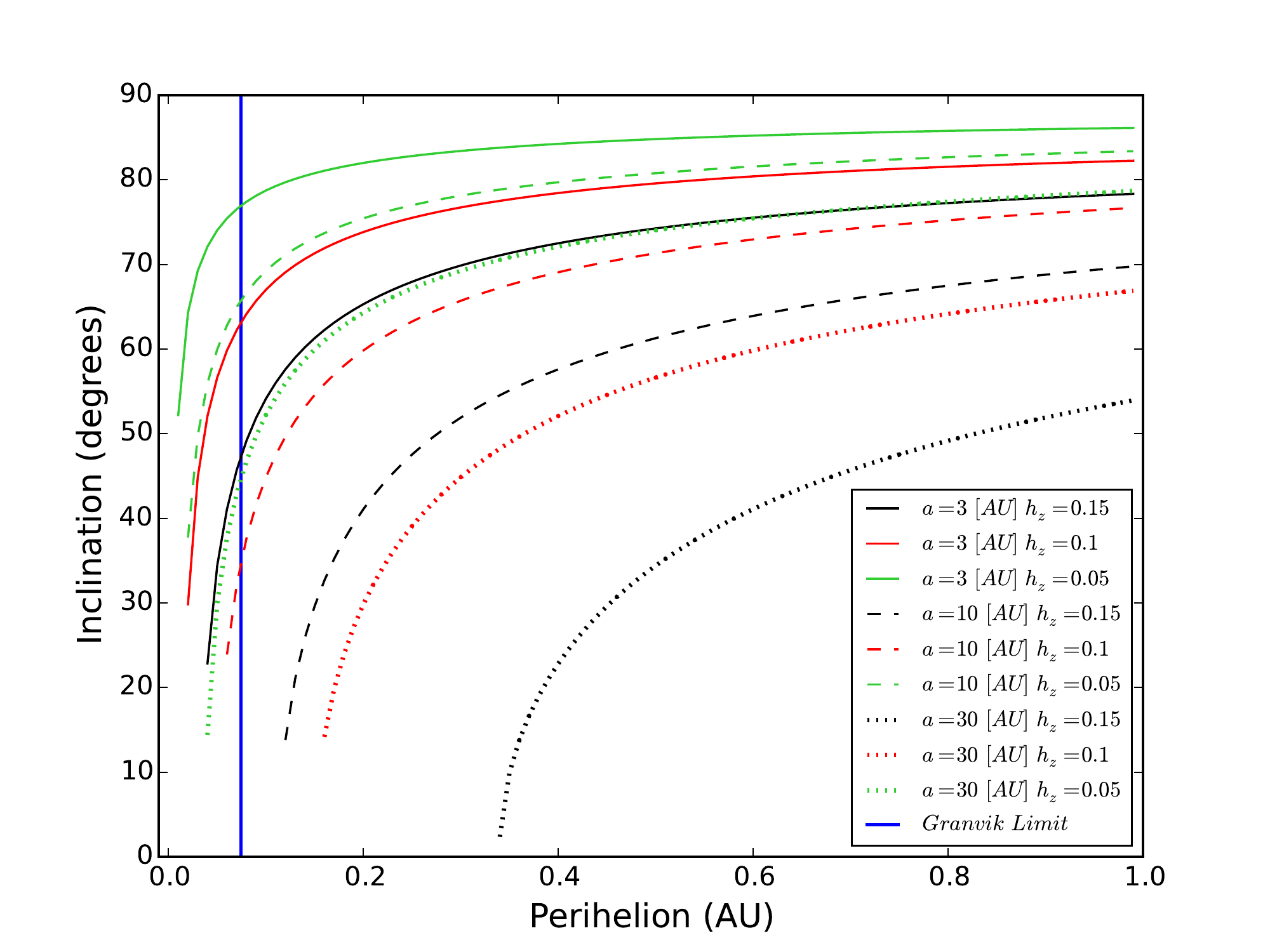}
\caption{The lines of constant $z$-component of the angular momentum $h_z= \sqrt{1-e^2} \cos i$, indicating the paths meteoroids follow under the Kozai effect. The vertical blue line indicates the G16 limit. Lines that cross this limit indicate the possible delivery of meteoroids to near-Sun space from high-inclination comets.  \label{fi:kozai}}
\end{center}
\end{figure}

Could this material be drawn into the near-Sun region by Kozai oscillations? These are caused by the long-term gravitational effects of the planets, and create correlated changes in $q$ and $i$ so that an orbit with larger $q$ and higher $i$ is drawn down to an orbit with smaller $q$ and lower $i$, the latter being rather like the meteoroid orbits in our sample. Figure~\ref{fi:kozai} shows the lines of constant $z$-component of the angular momentum, which meteoroids follow under the Kozai effect. We can see qualitatively that for dust to be emplaced inside the G16 limit by Kozai oscillations, the dust must 1) already be at low-perihelion distance at moderate ($30-60\degree$) inclination or 2) at high ($60-90\degree$) inclination without much constraint on $q$. We've already eliminated the known periodic low-perihelion comets as possible sources from our earlier analysis, so 1) is not an issue. High inclination periodic comets could supply this dust, and they are common so 2) is more difficult to eliminate as a possibility. However, high-inclination comets have orbits that are typically much larger than those of the dust observed by the meteor radar ($a \sim 2~$AU), though the orbit-shrinking effects of PR drag may account for some of the difference. 

So the two clearest possible sources for the dust in our sample are either relatively young dust from small SCDing NEAs/JFCs, these being drawn into the near-Sun region by the same process which drives near-Earth asteroids into the Sun \citep{farfrofro94}; or old dust from high-inclination comets which has been driven by Kozai oscillations into its current orbit. The model of \cite{pokvoknes14} indicates that relatively little dust from Halley-type comets makes it to low inclination low-perihelion orbits, but there is some. A careful modelling of dust orbital evolution under the effects of PR drag and the planets would be necessary to distinguish between SCD, cometary and Kozai-driven processes, but that beyond the scope of this paper. Nonetheless we conclude that there is near-Sun dust on orbits consistent with small SCDing asteroids, though such asteroids may not be the only possible source of such dust.

\section{Meteoroid erosion as the cause of SCD}
The cause of SCD remains unclear. G16 showed that tidal effects and evaporation could not account for the process, suggested that thermal cracking, spin-up and subsequent disruption or the heating of subsurface volatile pockets might provide a mechanism. Here we propose the SCD process is produced by erosion of them by near-Sun meteoroids. If this is the case, such erosion should affect both asteroids and comets in the near-Sun region, and we suggest that the anomalous brightening of near-Sun comets can also be explained by meteoroid erosion as well. Alternative explanations for the anomalous brightening of near-Sun comets includes the sublimation of olivine and pyroxene from dust grains \citep{kimmanbie02}, the onset sublimation in a previously thermally stable components of the nucleus \citep{kniahebie10}

\subsection{Phenomena which support meteoroid erosion as the cause of SCD}

\subsubsection{Anomalously fast near-Sun brightening}
The sunlight received by a body in orbit is proportional to one over the heliocentric distance $r_h$ squared. Near-Sun comets typically brighten faster than this, and faster than is typical of regular comets (which often go like $r_h^{-4}$, \cite{fesricwes93a}).  \cite{kniahebie10}'s analysis of nearly 1000 Kreutz comets observed by SOHO shows a stage of rapid brightening following $r_h^{-7.3\pm2.0}$ from an unknown heliocentric distance (but that it is ``unlikely'' to be beyond $50 R_{\odot}$) to $24 R_{\odot}$, where the brightening then drops to $r_h^{-3.8\pm 0.7}$.   \cite{huiyekni15} found $r_h^{-5.5}$ dependence for SOHO comet C/2015~D1 on its inbound leg; and \cite{yehuikra14} reported on 2 Kreutz comets whose brightening profiles had exponents consistent with -4, while 3 other comets had sharper values (-7 and beyond).

The kinetic energy flux of radially infalling meteoroids goes like $r^{-3.5}$, increasing more steeply than solar illumination and could better account for the observed rapid brightening.

\subsubsection{ The universality of Kreutz light curves}
\cite{bielamstc02} find that Kreutz sun-grazing ''comets all reach a peak brightness at one of two characteristic distances (both near 12 R$_{\odot}$) and that the comets fragment at another characteristic distance (about 7 R$_\odot$).'' They also say later ''The similarity of the light curves is remarkable in that all reach a peak brightness at about the same heliocentric distance. In addition, the shape of the curves are also very similar.''  Dimming after peak brightness is reached could be accounted for by volatile depletion perhaps,  but the uncorrelated spins, shapes, sizes, inhomogeneities of composition, etc. of these comets make universal light curves based solely on volatile release unlikely at best.

However, such behaviour could naturally be produced by the passage of the comets through a near-Sun meteoroid stream. If this stream was old enough to have dispersed around its orbit, all Kreutz comets passing through this portion of space would encounter a flux of high-speed particles that could release gas and dust in consistent, repeatable manner independent of the precise properties of the surface or subsurface volatile distribution of the comet itself.

Do non-Kreutz i.e sunskirters show similarly repeatable light curves? \cite{lamfaulle13} say that the light curves of the sun-skirters (Meyer; Marsden \& Kracht = Machholz and ungrouped) exhibit different behaviour but that "this primarily results from the differences in the distance ranges in which the measurements were made". So whether the light curves of non-Kreutz comets can be as easily attributed to a coherent near-Sun meteoroid environment is less clear. They do note the most common behaviour is "a continuous increase of the brightness as the comet approaches perihelion, reaching a peak before perihelion then progressively fading". This commonality is particularly intriguing since solar energy input must continue to increase as $r_h$ decreases, but the meteoroid flux need not.

\subsubsection{Brightening often diminishes despite moving closer to the Sun}
SOHO comets often diminish or remain constant in brightness despite continuing to approach the Sun, and this behaviour shows some correlation with the comet family. \cite{lamfaulle13} say "the real surprise comes from many comets (mostly from the Machholz group) with nearly 'flat' light curves.".  This could be accounted for by inhomogeneities in the meteoroid environment, such that the meteoroid flux closer to the Sun is less than further away, owing either to how the meteoroids were deposited by their parents, their subsequent dynamical evolution, or their destruction by thermal processes \citep{capvok10} or collisions with other meteoroids.

\subsection{The characteristics of meteoroid erosion}
 
Having proposed meteoroid erosion as the cause of SCD, how does it stack up to other destructive processes? The speed of a meteoroid on a nearly-circular orbit near the Sun is
\begin{equation}
v \approx 134 \left( \frac{r_h}{10~R_{\odot}} \right)^{-1/2} {\rm km/s}.
\end{equation}
At a distance of $r_h =10 R_\odot = 0.0465$ AU, a 6 mm diameter meteoroid (density 1000 kg~m$^{-3}$) on a retrograde circular orbit encountering a prograde asteroid deposits a kinetic energy of about 1 MJ, equal to that of a stick of dynamite.\footnote{https://www.chemistryviews.org/details/ezine/3622oid371/145\_Years\_of\_Dynamite.html} Retrograde meteor orbits (originating from Oort cloud comets) are abundant at the Earth and spiral into the Sun through PR drag, though meteoroids on less extreme orbits may provide erosive impacts of reduced but similar energies. There's no question that high-speed meteoroid impacts are physically able to remove material from and eventually destroy asteroids if present in sufficient numbers. But are there enough of them?

Unfortunately Earth-based meteor radars or cameras cannot measure the true near-Sun meteoroid population, but can only detect those that remain on highly-elongated orbits. However, the circularization and inspiralling of meteoroid orbits through Poynting-Robertson drag means that there is likely a substantial population of near-Sun meteoroids continually being fed into the near-Sun region. Dynamical models of these are starting to be constructed (e.g. \cite{poksarjan18} ) though these regions are largely inaccessible to meteor measurement techniques at Earth.

\cite{wie15} studied the effect of meteoroids as a source of drag on the asteroidal population near the Earth, and we extrapolate that model to the near-Sun region to assess the meteoroid impact rate on near-Sun objects. The flux of meteoroids onto an asteroid on a circular orbit at $r_h = 1$~AU was estimated at $3 \times 10^{-8}$~m$^{-2}$~s$^{-1}$ (about 1 impact per year per square meter) at a meteoroid mass of $1.5 \times 10^{-8}$~kg (300 $\mu$m diameter at a density of 1000~kg~m$^{-3}$). The dynamical evolution of such dust from the Earth to the near-Sun region is complex, but if we for simplicity assume it is mostly on nearly radial orbits (the near-Earth dust complex is known to be dominated by the helion and anti-helion sporadic meteor sources which contain just such radial orbits \citep{jonbro93}), then the number of impacts will scale like $r_h^{-2.5}$  ($-2$ for simple geometry with an additional 0.5 for the $r_h^{-0.5}$ speed dependence), or by a factor of $\approx 1800$ at $10 R_{\odot}$. The 300~$\mu$m meteoroid mentioned earlier carries a kinetic energy of 400~J at $10~R_{\odot}$, comparable to a bullet fired from a handgun, and our target asteroid is hit several times per day per square meter.

Whether this is sufficient to provide for the erosion necessary to produce the disruption of asteroids on its own is unclear but seems promising. In addition to direct removal, impacts could uncover fresher, more volatile-rich material: for example, there is evidence that lunar meteoroid impacts must exceed a certain energy before they can release water from the surface, owing to a desiccated layer several centimeters deep \citep{benhurstu19}. The speed and number of meteoroids near the Sun could cause the cracking of an object's low-thermal-conductivity mantle with subsequent burst of released volatiles.

In Section~\ref{ungroupedSOHO} it was noted that the fact that small SOHO comets don't survive even a single perihelion passage may require the removal of several meters of material from their surfaces during the few days spent in the near Sun region, and the impact rate computed above may not be sufficient to do so. The disruption of a stable low-thermal conductivity mantle resulting in subsurface volatile exposure to sunlight could be highly destructive, but the effects of high-speed impacts into loose regolith are not known.  However, given the order-of-magnitude of the effect of meteoroid erosion it seems likely an important contributor to the destruction of near-Sun asteroids. 

\section{Conclusions}

The process of disruption of near-Sun comets has been examined in the context of the known near-Earth asteroids, SOHO comets and meteoroid populations at different sizes. 

The past dynamical history of known NEAs is supportive of the SCD hypothesis in that there aren't many known NEAs which have spent time within the G16 limit. The exception is (467372) 2004 LG which spent 2500 years within that limit and survived, having now evolved to a larger perihelion.  Phaethon may represent a curious boundary case, in that its perihelion remains near but does not cross the G16 limit.

SOHO comets inform the study of SCD, but most SOHO comet families (Kreutz, Meyer and Machholz) are better described as originating from ordinary comet fragmentation than SCD. The exception are the ungrouped SOHO comets, which have an excess of orbits in the ecliptic plane consistent with the rate at which small NEAs are expected to be injected into this region. Thus these latter comets may in fact be asteroidal in nature, and additional study of this possibility is recommended.

There is an absence of meter-class fireballs seen with perihelia in the near-Sun region, but a population of millimeter-sized meteoroids (that may be but are not obviously cometary) extends well inside the G16 limit. This suggests that asteroids do not disrupt into meter-sized pieces but may break up into smaller pieces. The recovered meteorites with the two lowest known perihelia are relatively fragile, low bulk density, low albedo carbonaceous chondrites, though whether any of these features are clues to the SCD process or simply coincidence is not yet clear. 

We propose that the supercatastrophic disruption of near-Sun asteroids is due to meteoroid erosion. Though the population of near-Sun meteoroids is unknown, extrapolation from the Earth indicates that high-energy meteoroid impacts occur frequently on and deliver considerable energy to near-Sun bodies. Meteoroid impacts could also reproduce some of the puzzling brightening features seen in SOHO comets.  The supercatastrophic disruption of asteroids due to meteoroid impact could include direct removal of material by cratering ('sandblasting'), the exposure of sub-surface volatiles by mantle removal, and fracturing into several pieces.

Testing of the meteoroid erosion hypothesis and distinguishing it from competing processes will not be easy. The careful analysis of the size-distribution of near-Sun meteoroids is one possible approach, since the resultant size distribution is likely to be different for different processes. Other meteoroid impact/erosion processes in the solar system, such as the particle ejection events from asteroid (101955) Bennu \citep{lauherche19}, are also likely to provide valuable clues. Ultimately, reliable measurements and/or models of the near-Sun meteoroid environment will be needed to assess the true importance of meteoroid impacts on bodies travelling through this region.

\acknowledgements
We thank Bill Bottke for a thorough and thoughtful review that much improved this manuscript; Mikael Granvik and co-authors for providing the files associated with the \cite{gramorjed18} near-Earth asteroid model; and Rob Weryk for helpful discussions of the project and for locating Pan-STARRS images of asteroid (467372) 2004 LG. This research used the facilities of the Canadian Astronomy Data Centre operated by the National Research Council of Canada with the support of the Canadian Space Agency. PGB also acknowledges funding support from the Canada Research Chairs program. Funding for this work was provided through NASA co-operative agreement 80NSSC18M0046, the Natural Sciences and Engineering Research Council of Canada (Grants no. RGPIN-2016-04433 \& RGPIN-2018-05659) and the Canada Research Chairs Program and NASA ISFM award.


\bibliographystyle{model5-names}

\bibliography{Wiegert}

\begin{thebibliography}{73}
\expandafter\ifx\csname natexlab\endcsname\relax\def\natexlab#1{#1}\fi
\providecommand{\bibinfo}[2]{#2}
\ifx\xfnm\relax \def\xfnm[#1]{\unskip,\space#1}\fi
\bibitem[{{Abedin} et~al.(2018){Abedin}, {Wiegert}, {Janches}, {Pokorn{\'y}},
  {Brown} \& {Hormaechea}}]{abewiejan18}
\bibinfo{author}{{Abedin}, A.}, \bibinfo{author}{{Wiegert}, P.},
  \bibinfo{author}{{Janches}, D.}, \bibinfo{author}{{Pokorn{\'y}}, P.},
  \bibinfo{author}{{Brown}, P.}, \& \bibinfo{author}{{Hormaechea}, J.~L.}
  (\bibinfo{year}{2018}).
\newblock \bibinfo{title}{{Formation and past evolution of the showers of
  96P/Machholz complex}}.
\newblock {\it \bibinfo{journal}{icarus}\/},  {\it \bibinfo{volume}{300}\/},
  \bibinfo{pages}{360--385}.
\bibitem[{{Battams} \& {Knight}(2017)}]{batkni17}
\bibinfo{author}{{Battams}, K.}, \& \bibinfo{author}{{Knight}, M.~M.}
  (\bibinfo{year}{2017}).
\newblock \bibinfo{title}{{SOHO comets: 20 years and 3000 objects later}}.
\newblock {\it \bibinfo{journal}{Philosophical Transactions of the Royal
  Society of London Series A}\/},  {\it \bibinfo{volume}{375}\/},
  \bibinfo{pages}{20160257}.
\bibitem[{Beech \& Nikolova(1999)}]{Beech1999}
\bibinfo{author}{Beech, M.}, \& \bibinfo{author}{Nikolova, S.}
  (\bibinfo{year}{1999}).
\newblock \bibinfo{title}{{Large meteoroids in the Lyrid stream}}.
\newblock {\it \bibinfo{journal}{Mon. Not. Roy. Astron. Soc.}\/},  {\it
  \bibinfo{volume}{305}\/}, \bibinfo{pages}{253--258}.
\bibitem[{{Benna} et~al.(2019){Benna}, {Hurley}, {Stubbs}, {Mahaffy} \&
  {Elphic}}]{benhurstu19}
\bibinfo{author}{{Benna}, M.}, \bibinfo{author}{{Hurley}, D.~M.},
  \bibinfo{author}{{Stubbs}, T.~J.}, \bibinfo{author}{{Mahaffy}, P.~R.}, \&
  \bibinfo{author}{{Elphic}, R.~C.} (\bibinfo{year}{2019}).
\newblock \bibinfo{title}{{Lunar soil hydration constrained by exospheric water
  liberated by meteoroid impacts}}.
\newblock {\it \bibinfo{journal}{Nature Geoscience}\/},  {\it
  \bibinfo{volume}{12}\/}, \bibinfo{pages}{333--338}.
\bibitem[{{Biesecker} et~al.(2002){Biesecker}, {Lamy}, {St. Cyr}, {Llebaria} \&
  {Howard}}]{bielamstc02}
\bibinfo{author}{{Biesecker}, D.~A.}, \bibinfo{author}{{Lamy}, P.},
  \bibinfo{author}{{St. Cyr}, O.~C.}, \bibinfo{author}{{Llebaria}, A.}, \&
  \bibinfo{author}{{Howard}, R.~A.} (\bibinfo{year}{2002}).
\newblock \bibinfo{title}{{Sungrazing Comets Discovered with the SOHO/LASCO
  Coronagraphs 1996-1998}}.
\newblock {\it \bibinfo{journal}{Icarus}\/},  {\it \bibinfo{volume}{157}\/},
  \bibinfo{pages}{323--348}.
\bibitem[{Boenhardt(2004)}]{boe04}
\bibinfo{author}{Boenhardt, H.} (\bibinfo{year}{2004}).
\newblock \bibinfo{title}{Split comets}.
\newblock In \bibinfo{editor}{M.~Festou}, \bibinfo{editor}{H.~U. Keller}, \&
  \bibinfo{editor}{H.~A.~W. Jr.} (Eds.), {\it \bibinfo{booktitle}{Comets II}\/}
  (pp. \bibinfo{pages}{300--316}).
\newblock \bibinfo{address}{Tucson}: \bibinfo{publisher}{U. Arizona Press}.
\bibitem[{{Britt} et~al.(1992){Britt}, {Tholen}, {Bell} \&
  {Pieters}}]{brithobel92}
\bibinfo{author}{{Britt}, D.~T.}, \bibinfo{author}{{Tholen}, D.~J.},
  \bibinfo{author}{{Bell}, J.~F.}, \& \bibinfo{author}{{Pieters}, C.~M.}
  (\bibinfo{year}{1992}).
\newblock \bibinfo{title}{{Comparison of asteroid and meteorite spectra:
  Classification by principal component analysis}}.
\newblock {\it \bibinfo{journal}{icarus}\/},  {\it \bibinfo{volume}{99}\/},
  \bibinfo{pages}{153--166}.
\bibitem[{{Britt} et~al.(2002){Britt}, {Yeomans}, {Housen} \&
  {Consolmagno}}]{briyeohou02}
\bibinfo{author}{{Britt}, D.~T.}, \bibinfo{author}{{Yeomans}, D.},
  \bibinfo{author}{{Housen}, K.}, \& \bibinfo{author}{{Consolmagno}, G.}
  (\bibinfo{year}{2002}).
\newblock \bibinfo{title}{{Asteroid Density, Porosity, and Structure}}.
\newblock {\it \bibinfo{journal}{Asteroids III}\/},  (pp.
  \bibinfo{pages}{485--500}).
\bibitem[{{Brown} et~al.(2002){Brown}, {Spalding}, {ReVelle}, {Tagliaferri} \&
  {Worden}}]{brosparev02}
\bibinfo{author}{{Brown}, P.}, \bibinfo{author}{{Spalding}, R.~E.},
  \bibinfo{author}{{ReVelle}, D.~O.}, \bibinfo{author}{{Tagliaferri}, E.}, \&
  \bibinfo{author}{{Worden}, S.~P.} (\bibinfo{year}{2002}).
\newblock \bibinfo{title}{{The flux of small near-Earth objects colliding with
  the Earth}}.
\newblock {\it \bibinfo{journal}{Nature}\/},  {\it \bibinfo{volume}{420}\/},
  \bibinfo{pages}{294--296}.
\bibitem[{{Brown} et~al.(2016){Brown}, {Wiegert}, {Clark} \&
  {Tagliaferri}}]{browiecla16}
\bibinfo{author}{{Brown}, P.}, \bibinfo{author}{{Wiegert}, P.},
  \bibinfo{author}{{Clark}, D.}, \& \bibinfo{author}{{Tagliaferri}, E.}
  (\bibinfo{year}{2016}).
\newblock \bibinfo{title}{{Orbital and physical characteristics of meter-scale
  impactors from airburst observations}}.
\newblock {\it \bibinfo{journal}{Icarus}\/},  {\it \bibinfo{volume}{266}\/},
  \bibinfo{pages}{96--111}.
\bibitem[{{Brown} et~al.(2010){Brown}, {Wong}, {Weryk} \&
  {Wiegert}}]{browonwer10}
\bibinfo{author}{{Brown}, P.}, \bibinfo{author}{{Wong}, D.~K.},
  \bibinfo{author}{{Weryk}, R.~J.}, \& \bibinfo{author}{{Wiegert}, P.}
  (\bibinfo{year}{2010}).
\newblock \bibinfo{title}{{A meteoroid stream survey using the Canadian Meteor
  Orbit Radar. II: Identification of minor showers using a 3D wavelet
  transform}}.
\newblock {\it \bibinfo{journal}{Icarus}\/},  {\it \bibinfo{volume}{207}\/},
  \bibinfo{pages}{66--81}.
\bibitem[{{Brown} et~al.(2000){Brown}, {Hildebrand}, {Zolensky}, {Grady},
  {Clayton}, {Mayeda}, {Tagliaferri}, {Spalding}, {MacRae}, {Hoffman},
  {Mittlefehldt}, {Wacker}, {Bird}, {Campbell}, {Carpenter}, {Gingerich},
  {Glatiotis}, {Greiner}, {Mazur}, {McCausland}, {Plotkin} \& {Rubak
  Mazur}}]{brohilzol00}
\bibinfo{author}{{Brown}, P.~G.}, \bibinfo{author}{{Hildebrand}, A.~R.},
  \bibinfo{author}{{Zolensky}, M.~E.}, \bibinfo{author}{{Grady}, M.},
  \bibinfo{author}{{Clayton}, R.~N.}, \bibinfo{author}{{Mayeda}, T.~K.},
  \bibinfo{author}{{Tagliaferri}, E.}, \bibinfo{author}{{Spalding}, R.},
  \bibinfo{author}{{MacRae}, N.~D.}, \bibinfo{author}{{Hoffman}, E.~L.},
  \bibinfo{author}{{Mittlefehldt}, D.~W.}, \bibinfo{author}{{Wacker}, J.~F.},
  \bibinfo{author}{{Bird}, J.~A.}, \bibinfo{author}{{Campbell}, M.~D.},
  \bibinfo{author}{{Carpenter}, R.}, \bibinfo{author}{{Gingerich}, H.},
  \bibinfo{author}{{Glatiotis}, M.}, \bibinfo{author}{{Greiner}, E.},
  \bibinfo{author}{{Mazur}, M.~J.}, \bibinfo{author}{{McCausland}, P.~J.},
  \bibinfo{author}{{Plotkin}, H.}, \& \bibinfo{author}{{Rubak Mazur}, T.}
  (\bibinfo{year}{2000}).
\newblock \bibinfo{title}{{The Fall, Recovery, Orbit, and Composition of the
  Tagish Lake Meteorite: A New Type of Carbonaceous Chondrite}}.
\newblock {\it \bibinfo{journal}{Science}\/},  {\it \bibinfo{volume}{290}\/},
  \bibinfo{pages}{320--325}.
\bibitem[{Brown \& Jones(1995)}]{Brown1995}
\bibinfo{author}{Brown, P.~G.}, \& \bibinfo{author}{Jones, J.}
  (\bibinfo{year}{1995}).
\newblock \bibinfo{title}{{A determination of the strengths of the sporadic
  radio-meteor sources}}.
\newblock {\it \bibinfo{journal}{Earth, Moon, and Planets}\/},  {\it
  \bibinfo{volume}{68}\/}, \bibinfo{pages}{223--245}.
\bibitem[{Brown et~al.(2008)Brown, Weryk, Wong \& Jones}]{Brown2008}
\bibinfo{author}{Brown, P.~G.}, \bibinfo{author}{Weryk, R.},
  \bibinfo{author}{Wong, D.}, \& \bibinfo{author}{Jones, J.}
  (\bibinfo{year}{2008}).
\newblock \bibinfo{title}{{A meteoroid stream survey using the Canadian Meteor
  Orbit Radar I. Methodology and radiant catalogue}}.
\newblock {\it \bibinfo{journal}{Icarus}\/},  {\it \bibinfo{volume}{195}\/},
  \bibinfo{pages}{317--339}.
\bibitem[{{Chapman} et~al.(1975){Chapman}, {Morrison} \&
  {Zellner}}]{chamorzel75}
\bibinfo{author}{{Chapman}, C.~R.}, \bibinfo{author}{{Morrison}, D.}, \&
  \bibinfo{author}{{Zellner}, B.} (\bibinfo{year}{1975}).
\newblock \bibinfo{title}{{Surface Properties of Asteroids: A Synthesis of
  Polarimetry, Radiometry, and Spectrophotometry}}.
\newblock {\it \bibinfo{journal}{icarus}\/},  {\it \bibinfo{volume}{25}\/},
  \bibinfo{pages}{104--130}.
\bibitem[{Cotto-Figueroa et~al.(2016)Cotto-Figueroa, Asphaug, Garvie, Rai,
  Johnston, Borkowski, Datta, Chattopadhyay \& Morris}]{cotaspgar16}
\bibinfo{author}{Cotto-Figueroa, D.}, \bibinfo{author}{Asphaug, E.},
  \bibinfo{author}{Garvie, L.~A.}, \bibinfo{author}{Rai, A.},
  \bibinfo{author}{Johnston, J.}, \bibinfo{author}{Borkowski, L.},
  \bibinfo{author}{Datta, S.}, \bibinfo{author}{Chattopadhyay, A.}, \&
  \bibinfo{author}{Morris, M.~A.} (\bibinfo{year}{2016}).
\newblock \bibinfo{title}{Scale-dependent measurements of meteorite strength:
  Implications for asteroid fragmentation}.
\newblock {\it \bibinfo{journal}{Icarus}\/},  {\it \bibinfo{volume}{277}\/},
  \bibinfo{pages}{73 -- 77}.
\bibitem[{Devillepoix et~al.(2019)Devillepoix, Bland, Sansom, Towner,
  Cup{\'{a}}k, Howie, Hartig, Jansen-Sturgeon \& Cox}]{Devillepoix2019}
\bibinfo{author}{Devillepoix, H. A.~R.}, \bibinfo{author}{Bland, P.~A.},
  \bibinfo{author}{Sansom, E.~K.}, \bibinfo{author}{Towner, M.~C.},
  \bibinfo{author}{Cup{\'{a}}k, M.}, \bibinfo{author}{Howie, R.~M.},
  \bibinfo{author}{Hartig, B. A.~D.}, \bibinfo{author}{Jansen-Sturgeon, T.}, \&
  \bibinfo{author}{Cox, M.~A.} (\bibinfo{year}{2019}).
\newblock \bibinfo{title}{{Observation of metre-scale impactors by the Desert
  Fireball Network}}.
\newblock {\it \bibinfo{journal}{Monthly Notices of the Royal Astronomical
  Society}\/},  {\it \bibinfo{volume}{483}\/}, \bibinfo{pages}{5166--5178}.
\bibitem[{{Everhart}(1985)}]{eve85}
\bibinfo{author}{{Everhart}, E.} (\bibinfo{year}{1985}).
\newblock \bibinfo{title}{{An efficient integrator that uses Gauss-Radau
  spacings}}.
\newblock In \bibinfo{editor}{A.~{Carusi}}, \& \bibinfo{editor}{G.~B.
  {Valsecchi}} (Eds.), {\it \bibinfo{booktitle}{Dynamics of Comets: Their
  Origin and Evolution}\/} (pp. \bibinfo{pages}{185--202}).
\newblock \bibinfo{address}{Dordrecht}: \bibinfo{publisher}{Kluwer}.
\bibitem[{{Farinella} et~al.(1994){Farinella}, {Froeschl{\'e}},
  {Froeschl{\'e}}, {Gonczi}, {Hahn}, {Morbidelli} \& {Valsecchi}}]{farfrofro94}
\bibinfo{author}{{Farinella}, P.}, \bibinfo{author}{{Froeschl{\'e}}, C.},
  \bibinfo{author}{{Froeschl{\'e}}, C.}, \bibinfo{author}{{Gonczi}, R.},
  \bibinfo{author}{{Hahn}, G.}, \bibinfo{author}{{Morbidelli}, A.}, \&
  \bibinfo{author}{{Valsecchi}, G.~B.} (\bibinfo{year}{1994}).
\newblock \bibinfo{title}{{Asteroids falling into the Sun}}.
\newblock {\it \bibinfo{journal}{Nature}\/},  {\it \bibinfo{volume}{371}\/},
  \bibinfo{pages}{314--317}.
\bibitem[{Festou et~al.(1993a)Festou, Rickman \& West}]{fesricwes93a}
\bibinfo{author}{Festou, M.~C.}, \bibinfo{author}{Rickman, H.}, \&
  \bibinfo{author}{West, R.~M.} (\bibinfo{year}{1993a}).
\newblock \bibinfo{title}{Comets}.
\newblock {\it \bibinfo{journal}{Astron. Astrophys. Rev.}\/},  {\it
  \bibinfo{volume}{4}\/}, \bibinfo{pages}{363--447}.
\bibitem[{{Gladman} et~al.(1997){Gladman}, {Migliorini}, {Morbidelli},
  {Zappala}, {Michel}, {Cellino}, {Froeschle}, {Levison}, {Bailey} \&
  {Duncan}}]{glamigmor97}
\bibinfo{author}{{Gladman}, B.~J.}, \bibinfo{author}{{Migliorini}, F.},
  \bibinfo{author}{{Morbidelli}, A.}, \bibinfo{author}{{Zappala}, V.},
  \bibinfo{author}{{Michel}, P.}, \bibinfo{author}{{Cellino}, A.},
  \bibinfo{author}{{Froeschle}, C.}, \bibinfo{author}{{Levison}, H.~F.},
  \bibinfo{author}{{Bailey}, M.}, \& \bibinfo{author}{{Duncan}, M.}
  (\bibinfo{year}{1997}).
\newblock \bibinfo{title}{{Dynamical lifetimes of objects injected into
  asteroid belt resonances}}.
\newblock {\it \bibinfo{journal}{Science}\/},  {\it \bibinfo{volume}{277}\/},
  \bibinfo{pages}{197--201}.
\bibitem[{{Gounelle} et~al.(2006){Gounelle}, {Spurn{\'y}} \&
  {Bland}}]{gouspubla06}
\bibinfo{author}{{Gounelle}, M.}, \bibinfo{author}{{Spurn{\'y}}, P.}, \&
  \bibinfo{author}{{Bland}, P.~A.} (\bibinfo{year}{2006}).
\newblock \bibinfo{title}{{The orbit and atmospheric trajectory of the Orgueil
  meteorite from historical records}}.
\newblock {\it \bibinfo{journal}{Meteoritics and Planetary Science}\/},  {\it
  \bibinfo{volume}{41}\/}, \bibinfo{pages}{135--150}.
\bibitem[{{Granvik} et~al.(2016){Granvik}, {Morbidelli}, {Jedicke}, {Bolin},
  {Bottke}, {Beshore}, {Vokrouhlick{\'y}}, {Delb{\`o}} \&
  {Michel}}]{gramorjed16}
\bibinfo{author}{{Granvik}, M.}, \bibinfo{author}{{Morbidelli}, A.},
  \bibinfo{author}{{Jedicke}, R.}, \bibinfo{author}{{Bolin}, B.},
  \bibinfo{author}{{Bottke}, W.~F.}, \bibinfo{author}{{Beshore}, E.},
  \bibinfo{author}{{Vokrouhlick{\'y}}, D.}, \bibinfo{author}{{Delb{\`o}}, M.},
  \& \bibinfo{author}{{Michel}, P.} (\bibinfo{year}{2016}).
\newblock \bibinfo{title}{{Super-catastrophic disruption of asteroids at small
  perihelion distances}}.
\newblock {\it \bibinfo{journal}{Nature}\/},  {\it \bibinfo{volume}{530}\/},
  \bibinfo{pages}{303--306}.
\bibitem[{{Granvik} et~al.(2018){Granvik}, {Morbidelli}, {Jedicke}, {Bolin},
  {Bottke}, {Beshore}, {Vokrouhlick{\'y}}, {Nesvorn{\'y}} \&
  {Michel}}]{gramorjed18}
\bibinfo{author}{{Granvik}, M.}, \bibinfo{author}{{Morbidelli}, A.},
  \bibinfo{author}{{Jedicke}, R.}, \bibinfo{author}{{Bolin}, B.},
  \bibinfo{author}{{Bottke}, W.~F.}, \bibinfo{author}{{Beshore}, E.},
  \bibinfo{author}{{Vokrouhlick{\'y}}, D.}, \bibinfo{author}{{Nesvorn{\'y}},
  D.}, \& \bibinfo{author}{{Michel}, P.} (\bibinfo{year}{2018}).
\newblock \bibinfo{title}{{Debiased orbit and absolute-magnitude distributions
  for near-Earth objects}}.
\newblock {\it \bibinfo{journal}{Icarus}\/},  {\it \bibinfo{volume}{312}\/},
  \bibinfo{pages}{181--207}.
\bibitem[{{Greenstreet} et~al.(2012){Greenstreet}, {Ngo} \&
  {Gladman}}]{grengogla12}
\bibinfo{author}{{Greenstreet}, S.}, \bibinfo{author}{{Ngo}, H.}, \&
  \bibinfo{author}{{Gladman}, B.} (\bibinfo{year}{2012}).
\newblock \bibinfo{title}{{The orbital distribution of Near-Earth Objects
  inside Earth's orbit}}.
\newblock {\it \bibinfo{journal}{Icarus}\/},  {\it \bibinfo{volume}{217}\/},
  \bibinfo{pages}{355--366}.
\bibitem[{{Gwyn} et~al.(2012){Gwyn}, {Hill} \& {Kavelaars}}]{gwyhilkav12}
\bibinfo{author}{{Gwyn}, S.~D.~J.}, \bibinfo{author}{{Hill}, N.}, \&
  \bibinfo{author}{{Kavelaars}, J.~J.} (\bibinfo{year}{2012}).
\newblock \bibinfo{title}{{SSOS: A Moving-Object Image Search Tool for Asteroid
  Precovery}}.
\newblock {\it \bibinfo{journal}{Publ. Astr. Soc. Pacific}\/},  {\it
  \bibinfo{volume}{124}\/}, \bibinfo{pages}{579}.
\bibitem[{{Haack} et~al.(2012){Haack}, {Grau}, {Bischoff}, {Horstmann},
  {Wasson}, {S{\o}rensen}, {Laubenstein}, {Ott}, {Palme}, {Gellissen},
  {Greenwood}, {Pearson}, {Franchi}, {Gabelica} \&
  {Schmitt-Kopplin}}]{haagrabis12}
\bibinfo{author}{{Haack}, H.}, \bibinfo{author}{{Grau}, T.},
  \bibinfo{author}{{Bischoff}, A.}, \bibinfo{author}{{Horstmann}, M.},
  \bibinfo{author}{{Wasson}, J.}, \bibinfo{author}{{S{\o}rensen}, A.},
  \bibinfo{author}{{Laubenstein}, M.}, \bibinfo{author}{{Ott}, U.},
  \bibinfo{author}{{Palme}, H.}, \bibinfo{author}{{Gellissen}, M.},
  \bibinfo{author}{{Greenwood}, R.~C.}, \bibinfo{author}{{Pearson}, V.~K.},
  \bibinfo{author}{{Franchi}, I.~A.}, \bibinfo{author}{{Gabelica}, Z.}, \&
  \bibinfo{author}{{Schmitt-Kopplin}, P.} (\bibinfo{year}{2012}).
\newblock \bibinfo{title}{{Maribo{\textemdash}A new CM fall from Denmark}}.
\newblock {\it \bibinfo{journal}{Meteoritics and Planetary Science}\/},  {\it
  \bibinfo{volume}{47}\/}, \bibinfo{pages}{30--50}.
\bibitem[{{Halliday} \& {McIntosh}(1990)}]{halmci90}
\bibinfo{author}{{Halliday}, I.}, \& \bibinfo{author}{{McIntosh}, B.~A.}
  (\bibinfo{year}{1990}).
\newblock \bibinfo{title}{{Orbit of the Murchison Meteorite}}.
\newblock {\it \bibinfo{journal}{Meteoritics}\/},  {\it
  \bibinfo{volume}{25}\/}, \bibinfo{pages}{339}.
\bibitem[{{Hui} et~al.(2015){Hui}, {Ye}, {Knight}, {Battams} \&
  {Clark}}]{huiyekni15}
\bibinfo{author}{{Hui}, M.-T.}, \bibinfo{author}{{Ye}, Q.-Z.},
  \bibinfo{author}{{Knight}, M.}, \bibinfo{author}{{Battams}, K.}, \&
  \bibinfo{author}{{Clark}, D.} (\bibinfo{year}{2015}).
\newblock \bibinfo{title}{{Gone in a Blaze of Glory: The Demise of Comet C/2015
  D1 (SOHO)}}.
\newblock {\it \bibinfo{journal}{ApJ}\/},  {\it \bibinfo{volume}{813}\/},
  \bibinfo{pages}{73}.
\bibitem[{{Jenniskens}(2004)}]{jen04}
\bibinfo{author}{{Jenniskens}, P.} (\bibinfo{year}{2004}).
\newblock \bibinfo{title}{{2003 EH1 is the Quadrantid shower parent comet}}.
\newblock {\it \bibinfo{journal}{AJ}\/},  {\it \bibinfo{volume}{127}\/},
  \bibinfo{pages}{3018--3022}.
\bibitem[{{Jenniskens} et~al.(2012){Jenniskens}, {Fries}, {Yin}, {Zolensky},
  {Krot}, {Sandford}, {Sears}, {Beauford}, {Ebel}, {Friedrich}, {Nagashima},
  {Wimpenny}, {Yamakawa}, {Nishiizumi}, {Hamajima}, {Caffee}, {Welten},
  {Laubenstein}, {Davis}, {Simon}, {Heck}, {Young}, {Kohl}, {Thiemens}, {Nunn},
  {Mikouchi}, {Hagiya}, {Ohsumi}, {Cahill}, {Lawton}, {Barnes}, {Steele},
  {Rochette}, {Verosub}, {Gattacceca}, {Cooper}, {Glavin}, {Burton}, {Dworkin},
  {Elsila}, {Pizzarello}, {Ogliore}, {Schmitt-Kopplin}, {Harir}, {Hertkorn},
  {Verchovsky}, {Grady}, {Nagao}, {Okazaki}, {Takechi}, {Hiroi}, {Smith},
  {Silber}, {Brown}, {Albers}, {Klotz}, {Hankey}, {Matson}, {Fries}, {Walker},
  {Puchtel}, {Lee}, {Erdman}, {Eppich}, {Roeske}, {Gabelica}, {Lerche},
  {Nuevo}, {Girten} \& {Worden}}]{jenfriyin12}
\bibinfo{author}{{Jenniskens}, P.}, \bibinfo{author}{{Fries}, M.~D.},
  \bibinfo{author}{{Yin}, Q.-Z.}, \bibinfo{author}{{Zolensky}, M.},
  \bibinfo{author}{{Krot}, A.~N.}, \bibinfo{author}{{Sandford}, S.~A.},
  \bibinfo{author}{{Sears}, D.}, \bibinfo{author}{{Beauford}, R.},
  \bibinfo{author}{{Ebel}, D.~S.}, \bibinfo{author}{{Friedrich}, J.~M.},
  \bibinfo{author}{{Nagashima}, K.}, \bibinfo{author}{{Wimpenny}, J.},
  \bibinfo{author}{{Yamakawa}, A.}, \bibinfo{author}{{Nishiizumi}, K.},
  \bibinfo{author}{{Hamajima}, Y.}, \bibinfo{author}{{Caffee}, M.~W.},
  \bibinfo{author}{{Welten}, K.~C.}, \bibinfo{author}{{Laubenstein}, M.},
  \bibinfo{author}{{Davis}, A.~M.}, \bibinfo{author}{{Simon}, S.~B.},
  \bibinfo{author}{{Heck}, P.~R.}, \bibinfo{author}{{Young}, E.~D.},
  \bibinfo{author}{{Kohl}, I.~E.}, \bibinfo{author}{{Thiemens}, M.~H.},
  \bibinfo{author}{{Nunn}, M.~H.}, \bibinfo{author}{{Mikouchi}, T.},
  \bibinfo{author}{{Hagiya}, K.}, \bibinfo{author}{{Ohsumi}, K.},
  \bibinfo{author}{{Cahill}, T.~A.}, \bibinfo{author}{{Lawton}, J.~A.},
  \bibinfo{author}{{Barnes}, D.}, \bibinfo{author}{{Steele}, A.},
  \bibinfo{author}{{Rochette}, P.}, \bibinfo{author}{{Verosub}, K.~L.},
  \bibinfo{author}{{Gattacceca}, J.}, \bibinfo{author}{{Cooper}, G.},
  \bibinfo{author}{{Glavin}, D.~P.}, \bibinfo{author}{{Burton}, A.~S.},
  \bibinfo{author}{{Dworkin}, J.~P.}, \bibinfo{author}{{Elsila}, J.~E.},
  \bibinfo{author}{{Pizzarello}, S.}, \bibinfo{author}{{Ogliore}, R.},
  \bibinfo{author}{{Schmitt-Kopplin}, P.}, \bibinfo{author}{{Harir}, M.},
  \bibinfo{author}{{Hertkorn}, N.}, \bibinfo{author}{{Verchovsky}, A.},
  \bibinfo{author}{{Grady}, M.}, \bibinfo{author}{{Nagao}, K.},
  \bibinfo{author}{{Okazaki}, R.}, \bibinfo{author}{{Takechi}, H.},
  \bibinfo{author}{{Hiroi}, T.}, \bibinfo{author}{{Smith}, K.},
  \bibinfo{author}{{Silber}, E.~A.}, \bibinfo{author}{{Brown}, P.~G.},
  \bibinfo{author}{{Albers}, J.}, \bibinfo{author}{{Klotz}, D.},
  \bibinfo{author}{{Hankey}, M.}, \bibinfo{author}{{Matson}, R.},
  \bibinfo{author}{{Fries}, J.~A.}, \bibinfo{author}{{Walker}, R.~J.},
  \bibinfo{author}{{Puchtel}, I.}, \bibinfo{author}{{Lee}, C.-T.~A.},
  \bibinfo{author}{{Erdman}, M.~E.}, \bibinfo{author}{{Eppich}, G.~R.},
  \bibinfo{author}{{Roeske}, S.}, \bibinfo{author}{{Gabelica}, Z.},
  \bibinfo{author}{{Lerche}, M.}, \bibinfo{author}{{Nuevo}, M.},
  \bibinfo{author}{{Girten}, B.}, \& \bibinfo{author}{{Worden}, S.~P.}
  (\bibinfo{year}{2012}).
\newblock \bibinfo{title}{{Radar-Enabled Recovery of the
  Sutter{\textquoteright}s Mill Meteorite, a Carbonaceous Chondrite Regolith
  Breccia}}.
\newblock {\it \bibinfo{journal}{Science}\/},  {\it \bibinfo{volume}{338}\/},
  \bibinfo{pages}{1583}.
\bibitem[{{Jewitt} et~al.(2019){Jewitt}, {Asmus}, {Yang} \& {Li}}]{jewasmyan19}
\bibinfo{author}{{Jewitt}, D.}, \bibinfo{author}{{Asmus}, D.},
  \bibinfo{author}{{Yang}, B.}, \& \bibinfo{author}{{Li}, J.}
  (\bibinfo{year}{2019}).
\newblock \bibinfo{title}{{High-resolution Thermal Infrared Imaging of 3200
  Phaethon}}.
\newblock {\it \bibinfo{journal}{AJ}\/},  {\it \bibinfo{volume}{157}\/},
  \bibinfo{pages}{193}.
\bibitem[{{Jewitt} \& {Li}(2010)}]{jewli10}
\bibinfo{author}{{Jewitt}, D.}, \& \bibinfo{author}{{Li}, J.}
  (\bibinfo{year}{2010}).
\newblock \bibinfo{title}{{Activity in Geminid Parent (3200) Phaethon}}.
\newblock {\it \bibinfo{journal}{AJ}\/},  {\it \bibinfo{volume}{140}\/},
  \bibinfo{pages}{1519--1527}.
\bibitem[{{Jones} et~al.(2018){Jones}, {Knight}, {Battams}, {Boice}, {Brown},
  {Giordano}, {Raymond}, {Snodgrass}, {Steckloff}, {Weissman}, {Fitzsimmons},
  {Lisse}, {Opitom}, {Birkett}, {Bzowski}, {Decock}, {Mann}, {Ramanjooloo} \&
  {McCauley}}]{jonknibat18}
\bibinfo{author}{{Jones}, G.~H.}, \bibinfo{author}{{Knight}, M.~M.},
  \bibinfo{author}{{Battams}, K.}, \bibinfo{author}{{Boice}, D.~C.},
  \bibinfo{author}{{Brown}, J.}, \bibinfo{author}{{Giordano}, S.},
  \bibinfo{author}{{Raymond}, J.}, \bibinfo{author}{{Snodgrass}, C.},
  \bibinfo{author}{{Steckloff}, J.~K.}, \bibinfo{author}{{Weissman}, P.},
  \bibinfo{author}{{Fitzsimmons}, A.}, \bibinfo{author}{{Lisse}, C.},
  \bibinfo{author}{{Opitom}, C.}, \bibinfo{author}{{Birkett}, K.~S.},
  \bibinfo{author}{{Bzowski}, M.}, \bibinfo{author}{{Decock}, A.},
  \bibinfo{author}{{Mann}, I.}, \bibinfo{author}{{Ramanjooloo}, Y.}, \&
  \bibinfo{author}{{McCauley}, P.} (\bibinfo{year}{2018}).
\newblock \bibinfo{title}{{The Science of Sungrazers, Sunskirters, and Other
  Near-Sun Comets}}.
\newblock {\it \bibinfo{journal}{Space Sci. Rev.}\/},  {\it
  \bibinfo{volume}{214}\/}, \bibinfo{pages}{20}.
\bibitem[{{Jones} \& {Brown}(1993)}]{jonbro93}
\bibinfo{author}{{Jones}, J.}, \& \bibinfo{author}{{Brown}, P.}
  (\bibinfo{year}{1993}).
\newblock \bibinfo{title}{{Sporadic meteor radiant distributions - Orbital
  survey results}}.
\newblock {\it \bibinfo{journal}{MNRAS}\/},  {\it \bibinfo{volume}{265}\/},
  \bibinfo{pages}{524--532}.
\bibitem[{Jones et~al.(2005)Jones, Brown, Ellis, Webster, Campbell-Brown,
  Krzemenski \& Weryk}]{Jones2005}
\bibinfo{author}{Jones, J.}, \bibinfo{author}{Brown, P.~G.},
  \bibinfo{author}{Ellis, K.~J.}, \bibinfo{author}{Webster, A.},
  \bibinfo{author}{Campbell-Brown, M.~D.}, \bibinfo{author}{Krzemenski, Z.}, \&
  \bibinfo{author}{Weryk, R.} (\bibinfo{year}{2005}).
\newblock \bibinfo{title}{{The Canadian Meteor Orbit Radar : system overview
  and preliminary results}}.
\newblock {\it \bibinfo{journal}{Planetary and Space Science}\/},  {\it
  \bibinfo{volume}{53}\/}, \bibinfo{pages}{413--421}.
\bibitem[{Kaiser(1960)}]{Kaiser1960}
\bibinfo{author}{Kaiser, T.} (\bibinfo{year}{1960}).
\newblock \bibinfo{title}{{The determination of the incident flux of
  radio-meteors}}.
\newblock {\it \bibinfo{journal}{Monthly Notices of the Royal Astronomical
  Society}\/},  {\it \bibinfo{volume}{121}\/}, \bibinfo{pages}{284}.
\bibitem[{{Kimura} et~al.(2002){Kimura}, {Mann}, {Biesecker} \&
  {Jessberger}}]{kimmanbie02}
\bibinfo{author}{{Kimura}, H.}, \bibinfo{author}{{Mann}, I.},
  \bibinfo{author}{{Biesecker}, D.~A.}, \& \bibinfo{author}{{Jessberger},
  E.~K.} (\bibinfo{year}{2002}).
\newblock \bibinfo{title}{{Dust Grains in the Comae and Tails of Sungrazing
  Comets: Modeling of Their Mineralogical and Morphological Properties}}.
\newblock {\it \bibinfo{journal}{Icarus}\/},  {\it \bibinfo{volume}{159}\/},
  \bibinfo{pages}{529--541}.
\bibitem[{{Knight} et~al.(2010){Knight}, {A'Hearn}, {Biesecker}, {Faury},
  {Hamilton}, {Lamy} \& {Llebaria}}]{kniahebie10}
\bibinfo{author}{{Knight}, M.~M.}, \bibinfo{author}{{A'Hearn}, M.~F.},
  \bibinfo{author}{{Biesecker}, D.~A.}, \bibinfo{author}{{Faury}, G.},
  \bibinfo{author}{{Hamilton}, D.~P.}, \bibinfo{author}{{Lamy}, P.}, \&
  \bibinfo{author}{{Llebaria}, A.} (\bibinfo{year}{2010}).
\newblock \bibinfo{title}{{Photometric Study of the Kreutz Comets Observed by
  SOHO from 1996 to 2005}}.
\newblock {\it \bibinfo{journal}{AJ}\/},  {\it \bibinfo{volume}{139}\/},
  \bibinfo{pages}{926--949}.
\bibitem[{{Knight} et~al.(2016){Knight}, {Fitzsimmons}, {Kelley} \&
  {Snodgrass}}]{knifitkel16}
\bibinfo{author}{{Knight}, M.~M.}, \bibinfo{author}{{Fitzsimmons}, A.},
  \bibinfo{author}{{Kelley}, M. S.~P.}, \& \bibinfo{author}{{Snodgrass}, C.}
  (\bibinfo{year}{2016}).
\newblock \bibinfo{title}{{Comet 322P/SOHO 1: An Asteroid with the Smallest
  Perihelion Distance?}}
\newblock {\it \bibinfo{journal}{The Astrophysical Journal}\/},  {\it
  \bibinfo{volume}{823}\/}, \bibinfo{pages}{L6}.
\bibitem[{{Knight} \& {Walsh}(2013)}]{kniwal13}
\bibinfo{author}{{Knight}, M.~M.}, \& \bibinfo{author}{{Walsh}, K.~J.}
  (\bibinfo{year}{2013}).
\newblock \bibinfo{title}{{Will Comet ISON (C/2012 S1) Survive Perihelion?}}
\newblock {\it \bibinfo{journal}{apjl}\/},  {\it \bibinfo{volume}{776}\/},
  \bibinfo{pages}{L5}.
\bibitem[{{Kreutz}(1891)}]{kre91}
\bibinfo{author}{{Kreutz}, J.} (\bibinfo{year}{1891}).
\newblock {\it \bibinfo{journal}{Publ. Sternw. Kiel}\/}, .
\bibitem[{{Lamy} et~al.(2013){Lamy}, {Faury}, {Llebaria}, {Knight}, {A'Hearn}
  \& {Battams}}]{lamfaulle13}
\bibinfo{author}{{Lamy}, P.}, \bibinfo{author}{{Faury}, G.},
  \bibinfo{author}{{Llebaria}, A.}, \bibinfo{author}{{Knight}, M.~M.},
  \bibinfo{author}{{A'Hearn}, M.~F.}, \& \bibinfo{author}{{Battams}, K.}
  (\bibinfo{year}{2013}).
\newblock \bibinfo{title}{{Sunskirting comets discovered with the LASCO
  coronagraphs over the decade 1996-2008}}.
\newblock {\it \bibinfo{journal}{Icarus}\/},  {\it \bibinfo{volume}{226}\/},
  \bibinfo{pages}{1350--1398}.
\bibitem[{{Lamy} et~al.(2004){Lamy}, {Toth}, {Fernandez} \&
  {Weaver}}]{lamtotfer04}
\bibinfo{author}{{Lamy}, P.~L.}, \bibinfo{author}{{Toth}, I.},
  \bibinfo{author}{{Fernandez}, Y.~R.}, \& \bibinfo{author}{{Weaver}, H.~A.}
  (\bibinfo{year}{2004}).
\newblock \bibinfo{title}{{The sizes, shapes, albedos, and colors of cometary
  nuclei}}.
\newblock {\it \bibinfo{journal}{Comets II}\/},  (pp.
  \bibinfo{pages}{223--264}).
\bibitem[{{Lauretta} et~al.(2019){Lauretta}, {Hergenrother}, {Chesley},
  {Leonard}, {Pelgrift}, {Adam}, {Al Asad}, {Antreasian}, {Ballouz}, {Becker},
  {Bennett}, {Bos}, {Bottke}, {Brozovi{\'c}}, {Campins}, {Connolly}, {Daly},
  {Davis}, {de Le{\'o}n}, {DellaGiustina}, {Drouet d{\textquoteright}Aubigny},
  {Dworkin}, {Emery}, {Farnocchia}, {Glavin}, {Golish}, {Hartzell}, {Jacobson},
  {Jawin}, {Jenniskens}, {Kidd}, {Lessac-Chenen}, {Li}, {Libourel}, {Licand
  ro}, {Liounis}, {Maleszewski}, {Manzoni}, {May}, {McCarthy}, {McMahon},
  {Michel}, {Molaro}, {Moreau}, {Nelson}, {Owen}, {Rizk}, {Roper}, {Rozitis},
  {Sahr}, {Scheeres}, {Seabrook}, {Selznick}, {Takahashi}, {Thuillet},
  {Tricarico}, {Vokrouhlick{\'y}} \& {Wolner}}]{lauherche19}
\bibinfo{author}{{Lauretta}, D.~S.}, \bibinfo{author}{{Hergenrother}, C.~W.},
  \bibinfo{author}{{Chesley}, S.~R.}, \bibinfo{author}{{Leonard}, J.~M.},
  \bibinfo{author}{{Pelgrift}, J.~Y.}, \bibinfo{author}{{Adam}, C.~D.},
  \bibinfo{author}{{Al Asad}, M.}, \bibinfo{author}{{Antreasian}, P.~G.},
  \bibinfo{author}{{Ballouz}, R.~L.}, \bibinfo{author}{{Becker}, K.~J.},
  \bibinfo{author}{{Bennett}, C.~A.}, \bibinfo{author}{{Bos}, B.~J.},
  \bibinfo{author}{{Bottke}, W.~F.}, \bibinfo{author}{{Brozovi{\'c}}, M.},
  \bibinfo{author}{{Campins}, H.}, \bibinfo{author}{{Connolly}, H.~C.},
  \bibinfo{author}{{Daly}, M.~G.}, \bibinfo{author}{{Davis}, A.~B.},
  \bibinfo{author}{{de Le{\'o}n}, J.}, \bibinfo{author}{{DellaGiustina},
  D.~N.}, \bibinfo{author}{{Drouet d{\textquoteright}Aubigny}, C.~Y.},
  \bibinfo{author}{{Dworkin}, J.~P.}, \bibinfo{author}{{Emery}, J.~P.},
  \bibinfo{author}{{Farnocchia}, D.}, \bibinfo{author}{{Glavin}, D.~P.},
  \bibinfo{author}{{Golish}, D.~R.}, \bibinfo{author}{{Hartzell}, C.~M.},
  \bibinfo{author}{{Jacobson}, R.~A.}, \bibinfo{author}{{Jawin}, E.~R.},
  \bibinfo{author}{{Jenniskens}, P.}, \bibinfo{author}{{Kidd}, J.~N.},
  \bibinfo{author}{{Lessac-Chenen}, E.~J.}, \bibinfo{author}{{Li}, J.~Y.},
  \bibinfo{author}{{Libourel}, G.}, \bibinfo{author}{{Licand ro}, J.},
  \bibinfo{author}{{Liounis}, A.~J.}, \bibinfo{author}{{Maleszewski}, C.~K.},
  \bibinfo{author}{{Manzoni}, C.}, \bibinfo{author}{{May}, B.},
  \bibinfo{author}{{McCarthy}, L.~K.}, \bibinfo{author}{{McMahon}, J.~W.},
  \bibinfo{author}{{Michel}, P.}, \bibinfo{author}{{Molaro}, J.~L.},
  \bibinfo{author}{{Moreau}, M.~C.}, \bibinfo{author}{{Nelson}, D.~S.},
  \bibinfo{author}{{Owen}, W.~M.}, \bibinfo{author}{{Rizk}, B.},
  \bibinfo{author}{{Roper}, H.~L.}, \bibinfo{author}{{Rozitis}, B.},
  \bibinfo{author}{{Sahr}, E.~M.}, \bibinfo{author}{{Scheeres}, D.~J.},
  \bibinfo{author}{{Seabrook}, J.~A.}, \bibinfo{author}{{Selznick}, S.~H.},
  \bibinfo{author}{{Takahashi}, Y.}, \bibinfo{author}{{Thuillet}, F.},
  \bibinfo{author}{{Tricarico}, P.}, \bibinfo{author}{{Vokrouhlick{\'y}}, D.},
  \& \bibinfo{author}{{Wolner}, C.~W.~V.} (\bibinfo{year}{2019}).
\newblock \bibinfo{title}{{Episodes of particle ejection from the surface of
  the active asteroid (101955) Bennu}}.
\newblock {\it \bibinfo{journal}{Science}\/},  {\it \bibinfo{volume}{366}\/},
  \bibinfo{pages}{3544}.
\bibitem[{{Mainzer} et~al.(2011){Mainzer}, {Grav}, {Bauer}, {Masiero},
  {McMillan}, {Cutri}, {Walker}, {Wright}, {Eisenhardt}, {Tholen}, {Spahr},
  {Jedicke}, {Denneau}, {DeBaun}, {Elsbury}, {Gautier}, {Gomillion}, {Hand},
  {Mo}, {Watkins}, {Wilkins}, {Bryngelson}, {Del Pino Molina}, {Desai},
  {G{\'o}mez Camus}, {Hidalgo}, {Konstantopoulos}, {Larsen}, {Maleszewski},
  {Malkan}, {Mauduit}, {Mullan}, {Olszewski}, {Pforr}, {Saro}, {Scotti} \&
  {Wasserman}}]{maigrabau11}
\bibinfo{author}{{Mainzer}, A.}, \bibinfo{author}{{Grav}, T.},
  \bibinfo{author}{{Bauer}, J.}, \bibinfo{author}{{Masiero}, J.},
  \bibinfo{author}{{McMillan}, R.~S.}, \bibinfo{author}{{Cutri}, R.~M.},
  \bibinfo{author}{{Walker}, R.}, \bibinfo{author}{{Wright}, E.},
  \bibinfo{author}{{Eisenhardt}, P.}, \bibinfo{author}{{Tholen}, D.~J.},
  \bibinfo{author}{{Spahr}, T.}, \bibinfo{author}{{Jedicke}, R.},
  \bibinfo{author}{{Denneau}, L.}, \bibinfo{author}{{DeBaun}, E.},
  \bibinfo{author}{{Elsbury}, D.}, \bibinfo{author}{{Gautier}, T.},
  \bibinfo{author}{{Gomillion}, S.}, \bibinfo{author}{{Hand}, E.},
  \bibinfo{author}{{Mo}, W.}, \bibinfo{author}{{Watkins}, J.},
  \bibinfo{author}{{Wilkins}, A.}, \bibinfo{author}{{Bryngelson}, G.~L.},
  \bibinfo{author}{{Del Pino Molina}, A.}, \bibinfo{author}{{Desai}, S.},
  \bibinfo{author}{{G{\'o}mez Camus}, M.}, \bibinfo{author}{{Hidalgo}, S.~L.},
  \bibinfo{author}{{Konstantopoulos}, I.}, \bibinfo{author}{{Larsen}, J.~A.},
  \bibinfo{author}{{Maleszewski}, C.}, \bibinfo{author}{{Malkan}, M.~A.},
  \bibinfo{author}{{Mauduit}, J.~C.}, \bibinfo{author}{{Mullan}, B.~L.},
  \bibinfo{author}{{Olszewski}, E.~W.}, \bibinfo{author}{{Pforr}, J.},
  \bibinfo{author}{{Saro}, A.}, \bibinfo{author}{{Scotti}, J.~V.}, \&
  \bibinfo{author}{{Wasserman}, L.~H.} (\bibinfo{year}{2011}).
\newblock \bibinfo{title}{{NEOWISE Observations of Near-Earth Objects:
  Preliminary Results}}.
\newblock {\it \bibinfo{journal}{apj}\/},  {\it \bibinfo{volume}{743}\/},
  \bibinfo{pages}{156}.
\bibitem[{{Marsden}(1967)}]{mar67}
\bibinfo{author}{{Marsden}, B.~G.} (\bibinfo{year}{1967}).
\newblock \bibinfo{title}{{The sungrazing comet group}}.
\newblock {\it \bibinfo{journal}{AJ}\/},  {\it \bibinfo{volume}{72}\/},
  \bibinfo{pages}{1170}.
\bibitem[{{Masiero} et~al.(2011){Masiero}, {Mainzer}, {Grav}, {Bauer}, {Cutri},
  {Dailey}, {Eisenhardt}, {McMillan}, {Spahr}, {Skrutskie}, {Tholen}, {Walker},
  {Wright}, {DeBaun}, {Elsbury}, {Gautier}, {Gomillion} \&
  {Wilkins}}]{masmaigra11}
\bibinfo{author}{{Masiero}, J.~R.}, \bibinfo{author}{{Mainzer}, A.~K.},
  \bibinfo{author}{{Grav}, T.}, \bibinfo{author}{{Bauer}, J.~M.},
  \bibinfo{author}{{Cutri}, R.~M.}, \bibinfo{author}{{Dailey}, J.},
  \bibinfo{author}{{Eisenhardt}, P.~R.~M.}, \bibinfo{author}{{McMillan},
  R.~S.}, \bibinfo{author}{{Spahr}, T.~B.}, \bibinfo{author}{{Skrutskie},
  M.~F.}, \bibinfo{author}{{Tholen}, D.}, \bibinfo{author}{{Walker}, R.~G.},
  \bibinfo{author}{{Wright}, E.~L.}, \bibinfo{author}{{DeBaun}, E.},
  \bibinfo{author}{{Elsbury}, D.}, \bibinfo{author}{{Gautier}, I., T.},
  \bibinfo{author}{{Gomillion}, S.}, \& \bibinfo{author}{{Wilkins}, A.}
  (\bibinfo{year}{2011}).
\newblock \bibinfo{title}{{Main Belt Asteroids with WISE/NEOWISE. I.
  Preliminary Albedos and Diameters}}.
\newblock {\it \bibinfo{journal}{apj}\/},  {\it \bibinfo{volume}{741}\/},
  \bibinfo{pages}{68}.
\bibitem[{Mazur et~al.(2019)Mazur, Pokorn\'{y}, Weryk, Brown, Vida, Schult,
  Stober \& Agrawal}]{mazur2019}
\bibinfo{author}{Mazur, M.}, \bibinfo{author}{Pokorn\'{y}, P.},
  \bibinfo{author}{Weryk, R.~J.}, \bibinfo{author}{Brown, P.},
  \bibinfo{author}{Vida, D.}, \bibinfo{author}{Schult, C.},
  \bibinfo{author}{Stober, G.}, \& \bibinfo{author}{Agrawal, A.}
  (\bibinfo{year}{2019}).
\newblock \bibinfo{title}{An algorithm for measurement of radar transverse
  scattering meteor echo speeds using pre-$t_{0}$ phases}.
\newblock {\it \bibinfo{journal}{Radio Science}\/},  {\it
  \bibinfo{volume}{submitted}\/}.
\bibitem[{{Nishiizumi} et~al.(2014){Nishiizumi}, {Caffee}, {Hamajima}, {Reedy}
  \& {Welten}}]{niscafham14}
\bibinfo{author}{{Nishiizumi}, K.}, \bibinfo{author}{{Caffee}, M.~W.},
  \bibinfo{author}{{Hamajima}, Y.}, \bibinfo{author}{{Reedy}, R.~C.}, \&
  \bibinfo{author}{{Welten}, K.~C.} (\bibinfo{year}{2014}).
\newblock \bibinfo{title}{{Exposure history of the Sutter's Mill carbonaceous
  chondrite}}.
\newblock {\it \bibinfo{journal}{Meteoritics and Planetary Science}\/},  {\it
  \bibinfo{volume}{49}\/}, \bibinfo{pages}{2056--2063}.
\bibitem[{{{\"O}pik}(1951)}]{opi51}
\bibinfo{author}{{{\"O}pik}, E.} (\bibinfo{year}{1951}).
\newblock \bibinfo{title}{{Collision probability with the planets and the
  distribution of planetary matter}}.
\newblock {\it \bibinfo{journal}{Proc.~R.~Irish Acad.~Sect.~A, vol.~54,
  p.~165-199 (1951).}\/},  {\it \bibinfo{volume}{54}\/},
  \bibinfo{pages}{165--199}.
\bibitem[{{Pokorn{\'y}} et~al.(2018){Pokorn{\'y}}, {Sarantos} \&
  {Janches}}]{poksarjan18}
\bibinfo{author}{{Pokorn{\'y}}, P.}, \bibinfo{author}{{Sarantos}, M.}, \&
  \bibinfo{author}{{Janches}, D.} (\bibinfo{year}{2018}).
\newblock \bibinfo{title}{{A Comprehensive Model of the Meteoroid Environment
  around Mercury}}.
\newblock {\it \bibinfo{journal}{ApJ}\/},  {\it \bibinfo{volume}{863}\/},
  \bibinfo{pages}{31}.
\bibitem[{{Pokorn{\'y}} et~al.(2014){Pokorn{\'y}}, {Vokrouhlick{\'y}},
  {Nesvorn{\'y}}, {Campbell-Brown} \& {Brown}}]{pokvoknes14}
\bibinfo{author}{{Pokorn{\'y}}, P.}, \bibinfo{author}{{Vokrouhlick{\'y}}, D.},
  \bibinfo{author}{{Nesvorn{\'y}}, D.}, \bibinfo{author}{{Campbell-Brown}, M.},
  \& \bibinfo{author}{{Brown}, P.} (\bibinfo{year}{2014}).
\newblock \bibinfo{title}{{Dynamical Model for the Toroidal Sporadic Meteors}}.
\newblock {\it \bibinfo{journal}{ApJ}\/},  {\it \bibinfo{volume}{789}\/},
  \bibinfo{pages}{25}.
\bibitem[{{Scherer} \& {Schultz}(2000)}]{schsch00}
\bibinfo{author}{{Scherer}, P.}, \& \bibinfo{author}{{Schultz}, L.}
  (\bibinfo{year}{2000}).
\newblock \bibinfo{title}{{Noble gas record, collisional history and pairing of
  CV, CO, CK and other carbonaceous chondrites}}.
\newblock {\it \bibinfo{journal}{Meteoritics and Planetary Science}\/},  {\it
  \bibinfo{volume}{35}\/}, \bibinfo{pages}{145--153}.
\bibitem[{{Sekanina} \& {Chodas}(2005)}]{sekcho05}
\bibinfo{author}{{Sekanina}, Z.}, \& \bibinfo{author}{{Chodas}, P.~W.}
  (\bibinfo{year}{2005}).
\newblock \bibinfo{title}{{Origin of the Marsden and Kracht Groups of
  Sunskirting Comets. I. Association with Comet 96P/Machholz and Its
  Interplanetary Complex}}.
\newblock {\it \bibinfo{journal}{Astrophys. J. Suppl.}\/},  {\it
  \bibinfo{volume}{161}\/}, \bibinfo{pages}{551--586}.
\bibitem[{Spurny et~al.(2013)Spurny, Borovicka, Haack, Singer, Keuer \&
  Jobse}]{spuborhaa13}
\bibinfo{author}{Spurny, P.}, \bibinfo{author}{Borovicka, J.},
  \bibinfo{author}{Haack, H.}, \bibinfo{author}{Singer, W.},
  \bibinfo{author}{Keuer, D.}, \& \bibinfo{author}{Jobse, K.}
  (\bibinfo{year}{2013}).
\newblock \bibinfo{title}{{Trajectory and orbit of the Maribo CM2 Meteorite
  from optical, photoelectric and radar records}}.
\newblock In {\it \bibinfo{booktitle}{Meteoroids 2013}\/}.
\bibitem[{Spurn{\'{y}} et~al.(2017)Spurn{\'{y}}, Borovi{\v{c}}ka, Mucke \&
  Svoreň}]{Spurny2017}
\bibinfo{author}{Spurn{\'{y}}, P.}, \bibinfo{author}{Borovi{\v{c}}ka, J.},
  \bibinfo{author}{Mucke, H.}, \& \bibinfo{author}{Svoreň, J.}
  (\bibinfo{year}{2017}).
\newblock \bibinfo{title}{{Discovery of a new branch of the Taurid meteoroid
  stream as a real source of potentially hazardous bodies}}.
\newblock {\it \bibinfo{journal}{Astronomy {\&} Astrophysics}\/},  {\it
  \bibinfo{volume}{605}\/}, \bibinfo{pages}{A68}.
\bibitem[{{Tabeshian} et~al.(2019){Tabeshian}, {Wiegert}, {Ye}, {Hui}, {Gao} \&
  {Tan}}]{tabwieye19}
\bibinfo{author}{{Tabeshian}, M.}, \bibinfo{author}{{Wiegert}, P.},
  \bibinfo{author}{{Ye}, Q.}, \bibinfo{author}{{Hui}, M.-T.},
  \bibinfo{author}{{Gao}, X.}, \& \bibinfo{author}{{Tan}, H.}
  (\bibinfo{year}{2019}).
\newblock \bibinfo{title}{{Asteroid (3200) Phaethon: Colors, Phase Curve,
  Limits on Cometary Activity, and Fragmentation}}.
\newblock {\it \bibinfo{journal}{AJ}\/},  {\it \bibinfo{volume}{158}\/},
  \bibinfo{pages}{30}.
\bibitem[{{Tedesco} et~al.(1989){Tedesco}, {Williams}, {Matson}, {Weeder},
  {Gradie} \& {Lebofsky}}]{tedwilmat89}
\bibinfo{author}{{Tedesco}, E.~F.}, \bibinfo{author}{{Williams}, J.~G.},
  \bibinfo{author}{{Matson}, D.~L.}, \bibinfo{author}{{Weeder}, G.~J.},
  \bibinfo{author}{{Gradie}, J.~C.}, \& \bibinfo{author}{{Lebofsky}, L.~A.}
  (\bibinfo{year}{1989}).
\newblock \bibinfo{title}{{A Three-Parameter Asteroid Taxonomy}}.
\newblock {\it \bibinfo{journal}{aj}\/},  {\it \bibinfo{volume}{97}\/},
  \bibinfo{pages}{580}.
\bibitem[{{{\v{C}}apek} \& {Vokrouhlick{\'y}}(2010)}]{capvok10}
\bibinfo{author}{{{\v{C}}apek}, D.}, \& \bibinfo{author}{{Vokrouhlick{\'y}},
  D.} (\bibinfo{year}{2010}).
\newblock \bibinfo{title}{{Thermal stresses in small meteoroids}}.
\newblock {\it \bibinfo{journal}{A\&A}\/},  {\it \bibinfo{volume}{519}\/},
  \bibinfo{pages}{A75}.
\bibitem[{{Vokrouhlick{\'y}} \& {Nesvorn{\'y}}(2012)}]{voknes12}
\bibinfo{author}{{Vokrouhlick{\'y}}, D.}, \& \bibinfo{author}{{Nesvorn{\'y}},
  D.} (\bibinfo{year}{2012}).
\newblock \bibinfo{title}{{Sun-grazing orbit of the unusual near-Earth object
  2004 LG}}.
\newblock {\it \bibinfo{journal}{aap}\/},  {\it \bibinfo{volume}{541}\/},
  \bibinfo{pages}{A109}.
\bibitem[{Webster et~al.(2004)Webster, Brown, Jones, Ellis \&
  Campbell-Brown}]{webbrojon04}
\bibinfo{author}{Webster, A.~R.}, \bibinfo{author}{Brown, P.~G.},
  \bibinfo{author}{Jones, J.}, \bibinfo{author}{Ellis, K.~J.}, \&
  \bibinfo{author}{Campbell-Brown, M.} (\bibinfo{year}{2004}).
\newblock \bibinfo{title}{{The Canadian Meteor Orbit Radar (CMOR)}}.
\newblock {\it \bibinfo{journal}{Atmos. Chem. Phys.}\/},  {\it
  \bibinfo{volume}{4}\/}, \bibinfo{pages}{1181--1201}.
\bibitem[{{Whipple}(1940)}]{whi40}
\bibinfo{author}{{Whipple}, F.~L.} (\bibinfo{year}{1940}).
\newblock \bibinfo{title}{{Photographics meteor studies. III. The Taurid
  shower}}.
\newblock {\it \bibinfo{journal}{Proc. Amer. Phil. Soc.}\/},  {\it
  \bibinfo{volume}{83}\/}, \bibinfo{pages}{711--745}.
\bibitem[{{Whipple}(1951)}]{whi51}
\bibinfo{author}{{Whipple}, F.~L.} (\bibinfo{year}{1951}).
\newblock \bibinfo{title}{{A comet model. II. Physical relations for comets and
  meteors.}}
\newblock {\it \bibinfo{journal}{ApJ}\/},  {\it \bibinfo{volume}{113}\/},
  \bibinfo{pages}{464--474}.
\bibitem[{{Whipple}(1983)}]{whi83}
\bibinfo{author}{{Whipple}, F.~L.} (\bibinfo{year}{1983}).
\newblock \bibinfo{title}{{1983 TB and the Geminid Meteors}}.
\newblock {\it \bibinfo{journal}{IAU Circular}\/},  {\it
  \bibinfo{volume}{3881}\/}, \bibinfo{pages}{1}.
\bibitem[{Whipple \& Huebner(1976)}]{whihue76}
\bibinfo{author}{Whipple, F.~L.}, \& \bibinfo{author}{Huebner, W.~F.}
  (\bibinfo{year}{1976}).
\newblock \bibinfo{title}{Physical processes in comets}.
\newblock {\it \bibinfo{journal}{Ann. Rev. Astron. Astrophys.}\/},  {\it
  \bibinfo{volume}{14}\/}, \bibinfo{pages}{143--172}.
\bibitem[{{Wiegert} \& {Brown}(2005)}]{wiebro05}
\bibinfo{author}{{Wiegert}, P.}, \& \bibinfo{author}{{Brown}, P.}
  (\bibinfo{year}{2005}).
\newblock \bibinfo{title}{{The Quadrantid meteoroid complex}}.
\newblock {\it \bibinfo{journal}{Icarus}\/},  {\it \bibinfo{volume}{179}\/},
  \bibinfo{pages}{139--157}.
\bibitem[{{Wiegert}(2015)}]{wie15}
\bibinfo{author}{{Wiegert}, P.~A.} (\bibinfo{year}{2015}).
\newblock \bibinfo{title}{{Meteoroid impacts onto asteroids: A competitor for
  Yarkovsky and YORP}}.
\newblock {\it \bibinfo{journal}{Icarus}\/},  {\it \bibinfo{volume}{252}\/},
  \bibinfo{pages}{22--31}.
\bibitem[{{Wyatt} \& {Whipple}(1950)}]{wyawhi50}
\bibinfo{author}{{Wyatt}, S.~P.}, \& \bibinfo{author}{{Whipple}, F.~L.}
  (\bibinfo{year}{1950}).
\newblock \bibinfo{title}{{The Poynting-Robertson effect on meteor orbits}}.
\newblock {\it \bibinfo{journal}{ApJ}\/},  {\it \bibinfo{volume}{111}\/},
  \bibinfo{pages}{134--141}.
\bibitem[{Ye et~al.(2016)Ye, Brown \& Pokorn{\'{y}}}]{Ye2016}
\bibinfo{author}{Ye, Q.}, \bibinfo{author}{Brown, P.~G.}, \&
  \bibinfo{author}{Pokorn{\'{y}}, P.} (\bibinfo{year}{2016}).
\newblock \bibinfo{title}{{Dormant Comets Among the Near-Earth Object
  Population: A Meteor-Based Survey}}.
\newblock {\it \bibinfo{journal}{Mon. Not. R. Astron. Soc}\/},  {\it
  \bibinfo{volume}{462}\/}, \bibinfo{pages}{3511--3527}.
\bibitem[{{Ye} \& {Granvik}(2019)}]{yegra19}
\bibinfo{author}{{Ye}, Q.}, \& \bibinfo{author}{{Granvik}, M.}
  (\bibinfo{year}{2019}).
\newblock \bibinfo{title}{{Debris of Asteroid Disruptions Close to the Sun}}.
\newblock {\it \bibinfo{journal}{ApJ}\/},  {\it \bibinfo{volume}{873}\/},
  \bibinfo{pages}{104}.
\bibitem[{{Ye} et~al.(2018){Ye}, {Wiegert} \& {Hui}}]{yewiehui19}
\bibinfo{author}{{Ye}, Q.}, \bibinfo{author}{{Wiegert}, P.~A.}, \&
  \bibinfo{author}{{Hui}, M.-T.} (\bibinfo{year}{2018}).
\newblock \bibinfo{title}{{In Search of Recent Disruption of (3200) Phaethon:
  Model Implication and Hubble Space Telescope Search}}.
\newblock {\it \bibinfo{journal}{Astrophys. J. Lett.}\/},  {\it
  \bibinfo{volume}{864}\/}, \bibinfo{pages}{L9}.
\bibitem[{{Ye} et~al.(2014){Ye}, {Hui}, {Kracht} \& {Wiegert}}]{yehuikra14}
\bibinfo{author}{{Ye}, Q.-Z.}, \bibinfo{author}{{Hui}, M.-T.},
  \bibinfo{author}{{Kracht}, R.}, \& \bibinfo{author}{{Wiegert}, P.~A.}
  (\bibinfo{year}{2014}).
\newblock \bibinfo{title}{{Where are the Mini Kreutz-family Comets?}}
\newblock {\it \bibinfo{journal}{ApJ}\/},  {\it \bibinfo{volume}{796}\/},
  \bibinfo{pages}{83}.

\end{thebibliography}

\end{document}